\documentclass[11pt,a4paper]{article}
\usepackage[utf8]{inputenc}

\usepackage{jheppub}
\usepackage{amsthm,amsbsy,amsfonts,mathrsfs,enumerate,float,wrapfig,amsmath}
\usepackage[utf8]{inputenc}
\usepackage{subfigure}

\usepackage{tikz}
\usepackage{amsmath}
\usepackage{tikz-cd}
\usepackage{textcomp}
\usetikzlibrary{positioning}
\usetikzlibrary{arrows}

\makeatletter
\newcommand\xleftrightarrow[2][]{%
	\ext@arrow 9999{\longleftrightarrowfill@}{#1}{#2}}
\newcommand\longleftrightarrowfill@{%
	\arrowfill@\leftarrow\relbar\rightarrow}
\makeatother

\usepackage{extarrows}

\newcommand{\nn}{\nonumber}
\def\a{{\alpha}}

\def\b{{\beta}}

\def\0{{\emptyset}}

\def\inf{{\infty}}

\def\C{{\mathcal{C}}}
\def\N{{\mathcal{N}}}

\def\>{{  \succcurlyeq} }
\newcommand{\bsmall}{\begin{small}   }
	\newcommand{\esmall}{ \end{small}   }

\def\PE{\mathrm{PE}}

\def\N{ \mathcal{N}}
\def\C{ \mathbf{C}}

\def\T{ \mathcal{T}}
\def\Li{ \text{Li}  }
\def\log{ \text{log}}
\def\Weff{     \widetilde{\mathcal{W}}^{eff}      }
\def\W{ \mathcal{W}  }
\def\H{\mathcal{H}}
 \def\txi{\tilde\xi}
 \def\C{\mathbf{C}}
 
 \def\1{ \mathbf{1}}
  \def\0{ \mathbf{0}}
   \def\2{ \mathbf{2}}

\title{
Mirror Symmetry and Mixed Chern-Simons Levels for Abelian 3d $\N=2$ theories
}
\author[a]{Shi Cheng}

\affiliation[a]{Faculty of Physics, University of Warsaw, ul. Pasteura 5, 02-093 Warsaw, Poland}

\emailAdd{scheng@fuw.edu.pl}

\abstract{
We study the mirror symmetry of abelian 3d $\mathcal{N}=2$ theories with mixed Chern-Simons levels by turning them into $\T_{A,N}$ theories that are defined as $N$ copies of $U(1)-[1]$ theory coupled together by mixed Chern-Simons levels $k_{ij}$.
We find that $\T_{A,N}$ theories have many mirror dual theories with different mixed CS levels and FI parameters.
As an example,
we analyze $U(1)_k+ N_C \,\mathbf{C} + N_{AC} \, \mathbf{AC}$ theories by transforming these theories into certain
$\T_{A,N}$ theories and find many equivalent effective Chern-Simons levels. Finally, we analyze mirror symmetry for theories corresponding to knots. 
In this work we use sphere partition functions and vortex partition functions to derive dual theories.

}

\begin{document}
\maketitle
%=================================================================

%====================================================================
\section{Introduction}

Mirror symmetry relates many aspects of 3d $\N=2$ gauge theories, such as Seiberg dualities,  brane constructions, 3d/3d correspondence, etc., see \cite{Aharony:1997bx, Dimofte:2010tz, Dimofte:2011ju,Terashima:2011qi}. Constructing mirror pairs is a difficult task even for abelian theories.  Fortunately,  Kapustin and  Strassler found  in \cite{Kapustin:1999ha} that 3d mirror symmetry acts as functional Fourier transformation on partition functions, which provides an easy way to analyze 3d $\N=2$ gauge theories and  construct mirror dual theories, see e.g. \cite{Benvenuti:2016wet}. One subtle problem in mirror symmetry involves mixed Chern-Simons levels in 3d $\N=2$ theories, which have appeared e.g. in \cite{Beem:2012mb,Closset:2017zgf,Eckhard:2019aa}, but  have not been extensively studied yet. 
In addition, the recently discovered knots-quivers correspondence (KQ)
 implies that colored HOMFLY-PT polynomials for knots correspond to vortex partition functions of certain 3d $\N=2$ abelian theories with symmetric integer mixed Chern-Simons levels \cite{Ekholm:2018eee}. This motivates us to consider the physical interpretation of KQ correspondence and its relation to 3d quiver theories with mixed CS levels.

The 3d $\N=2$ mirror symmetry is naturally one important part of this story,  as it provides a powerful way to construct mirror dual pairs.  In order to consider mirror symmetry for theories with mixed CS levels we define a class of theories denoted by $\T_{A,N}$, which consist of a bunch of $U(1)-[1]$ theories coupled together by mixed Chern-Simons levels. We usually denote $\T_{A,N}$ theories by $(U(1)-[1])^{N}_{k_{ij}}$. The building block $U(1)-[1]$ of these theories is a theory that has one gauge group $U(1)$ and one chiral multiplet with charge $+1$. 
Moreover, it is found by Kapustin and Strassler in \cite{Kapustin:1999ha} that $U(1)-[1]$ is mirror to a free chiral multiplet denoted by $[1]-[1]$, and vice versa. Based on this, we find that the mirror symmetries (also called mirror transformations) acting on various building blocks commute with each other.  Altogether they form a nice mirror transformation group $\H(\T_{A,N})$.  For simplicity, we mainly discuss the mirror transformations of sphere partition functions, which at semi-classical limit give rise to effective superpotentials that encode  CS levels and FI parameters and label the 3d theories, and then verify the results by analysis of vortex partition functions. Since there are many mirror symmetries in $\H(\T_{A,N})$ and each of them gives rise to a mirror dual theory, it seems that we end up with many different mirror dual theories. 
However, these mirror dual theories are
equivalent and their partition functions are equal. %One can say these different CS levels obtained by mirror transformations are equivalent. 
%Therefore, mirror transformation provides a powerful way to find equivalent mixed CS levels. 
In addition, we need to take into account the parity anomaly constraints, which requires effective CS levels to be integers; hence only a subset of these mirror dual theories are consistent.   

To see the application of $\T_{A,N}$ theories, we discuss $U(1)-[N]$ theories, which have brane constructions dual to  strip Calabi-Yau threefolds with one open topological brane.
% and the associated vortex partition functions can be computed by topological strings methods. 
By applying mirror transformations on each chiral multiplet of $U(1)-[N]$, one can turn these theories into certain $\T_{A,N}$ theories, as illustrated in the following diagram
\begin{align}
	\centering
	\includegraphics[width=3in]{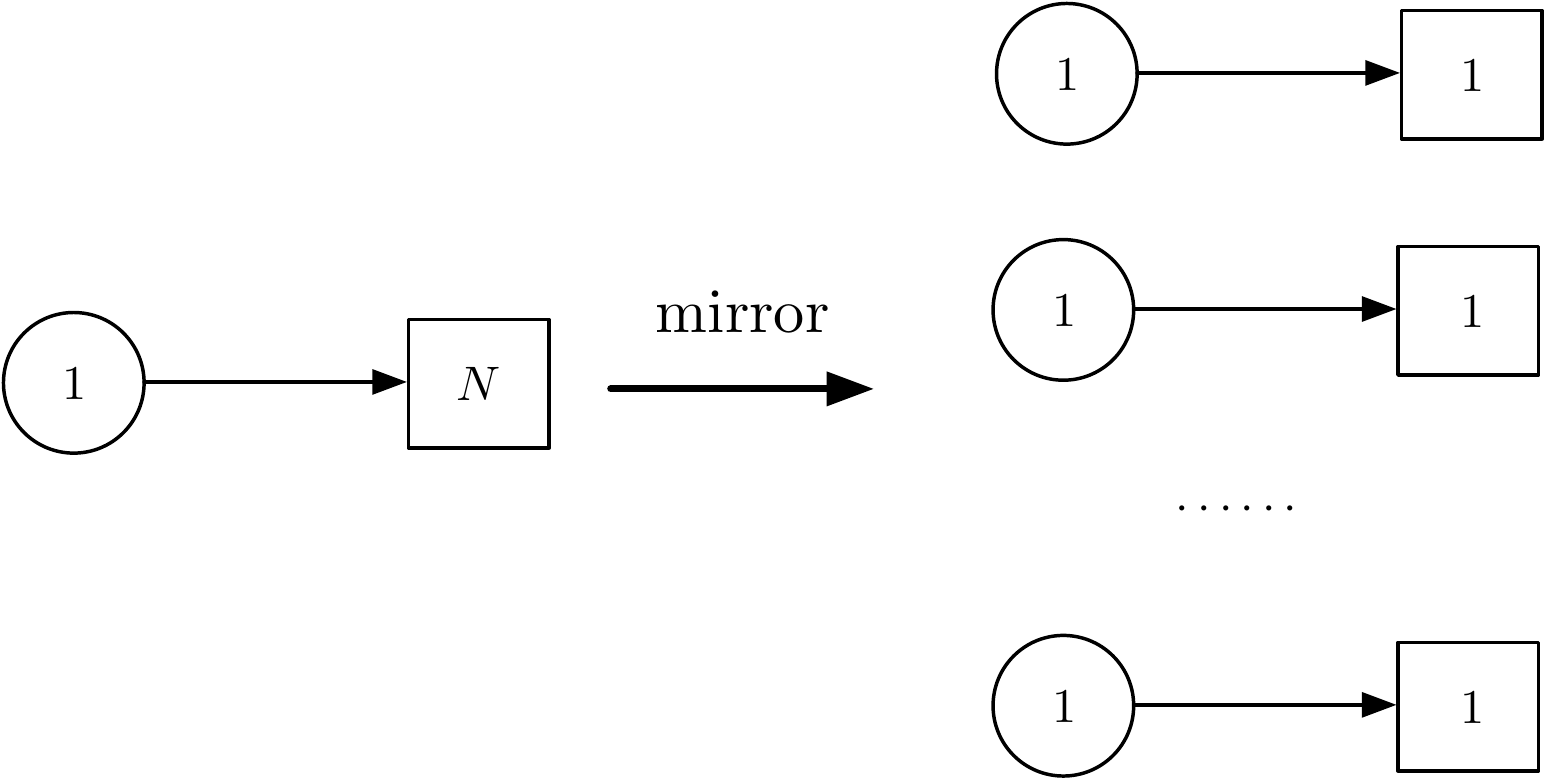} \,. \nn
\end{align}
This gives an easy way to perform mirror transformations on $U(1)-[N]$ theories.
Interestingly, we find the vortex partition functions of $U(1)-[N]$ can be written in the form of vortex partition functions of $\T_{A,N}$ theories, from which effective CS level matrices can be obtained by taking semi-classical limit. These mixed CS levels are the same as what we obtain from
%by performing mirror transformations on 
sphere partition functions.  In this example we find that mirror symmetry only flips the signs of mass parameters. 
%This confirm our statement that provides examples showing how mirror symmetry acts on vortex partition functions.

The paper is organized as follows.
In section \ref{section-2}, we review the localization method for 3d $\N=2$ theories, and show how mirror transformations act on sphere partition functions. The effective superpotentials and open Gopakumar-Vafa formula for 3d $\N=2$ theories are also discussed.
In section \ref{section-3}, 
we apply mirror symmetry on  theories engineered by strip Calabi-Yau threefolds by transforming them into  $\T_{A,N}$ theories. We also verify the diversity of mixed CS levels by analyzing vortex partition functions.
In section \ref{section-4}, we discuss the application of mirror symmetry on knot polynomials. 
Section \ref{sec-5} contains conclusions and a list of open problems.

\section{3d $\N=2$ mirror symmetry and $\T_{A,N}$ theory  }\label{section-2}
\subsection{Sphere partition function}

It is well known that localization techniques reduce the path integral representation of partition functions to finite dimensional contour integrals. 
%In this work we consider the 3d $\N=2$ gauge theories on three sphere $S_b^3$
In \cite{Hama:2011ea, Kapustin:2009kz}, the localization of 3d $\N=2$ gauge theories on three sphere 
\begin{align}
S_b^3: b^2 |z_1|^2 + b^{-2} |z_2|^2=1 \,, ~~z_1, z_2 \in \mathbb{C}
\end{align}
% (where $z_1, z_2 \in \mathbb{C}$)
  is developed, which shows that on Coulomb branch sphere partition functions can be written in terms of the contour integral of one-loop contributions from chiral multiplets and vector multiples. More explicitly,
the contribution from bare Chern-Simons level $k$ and FI term $\xi$ is
\begin{align}
\exp \big( - i\, \pi  \,  k x^2  + 2 \, i\, \pi \xi x  \big)
\,,
\end{align}
where $x$ is gauge transformation parameter for gauge group $U(1)_k$. 
The one-loop contributions from the fundamental chiral multiplet $\mathbf{C}$ and antifundamental chiral multiplet $\mathbf{AC}$ are
\begin{align}
s_b\Big( x+\frac{  i\, Q} {2}   + \frac{u}{2}  \Big)   \,, \quad  \frac{1}{s_b\Big( x-\frac{  i\, Q} {2}  - \frac{u}{2}  \Big) }  ,
\end{align}
respectively, where $Q=b+1/b$ is the localization parameter and $u$ is  the real mass parameter. 
%The chirals and anti-chirals are named because their charges are $1$ and $-1$ under gauge group. 
The contributions from antifundamental chiral multiplets can be written as 
\begin{align}
\frac{1}{s_b\Big( x-\frac{  i\, Q} {2}  - \frac{u}{2}  \Big) }  =s_b\Big( \frac{  i\, Q} {2}   -x+ \frac{u}{2}  \Big) \,.
\end{align}
%because of the identity for double sine function $s_b(x)s_b(-x)=1$,

For illustration, consider 3d $\N=2$ theories $U(1)_k+ N_{C} \mathbf{C}$. These theories have gauge group $U(1)$, bare Chern-Simons level $k$ and $N_C$ chiral multiplets $\mathbf{C}$. We denote them by quivers	
$(1)_k-[N_C]  $, and their sphere partition functions take the form
\begin{align}\label{shperepartNC}
Z_{S_b^3}^{ \,(1)_k-[N]\,}  = \int dx \,  e^{ - i\, \pi  \,  k x^2  + 2 \, i\, \pi \xi x  }  \prod\limits_{ i=1}^{N_C} s_b\Big( \frac{  i\, Q} {2}  +x + \frac{u_i}{2}  \Big)  \,.
\end{align}
Similarly, for theories $U(1)_k+ N_{C}\,\mathbf{C}+ N_{AC} \,\mathbf{AC} $, sphere partition functions take the form
\begin{align}\label{sphereNcNac}
Z_{S_b^3}^{U(1)_k+ N_{C}\,\mathbf{C}+ N_{AC} \,\mathbf{AC} }  = \int dx \,  e^{ - i\, \pi  \,  k x^2  + 2 \, i\, \pi \xi x  }  \prod\limits_{ i=1}^{N_C} s_b\Big( \frac{  i\, Q} {2}  +x + \frac{u_i}{2}  \Big) 
\prod\limits_{ j=1}^{N_{AC}} s_b\Big( \frac{  i\, Q} {2}  -x + \frac{u_j}{2}  \Big) 
\,.
\end{align}

In this work, we mainly consider the abelian quiver theories:
\begin{align}\label{quiverkij} 
\T_{A,N}:~~(U(1)-[1])^{  \otimes N}_{k_{ij},~\xi_i} \,,
\end{align}
which are $N$ copies of $U(1)-[1]$ theory,  with real symmetric Chern-Simons levels $k_{ij}$ between gauge groups $U(1)\times U(1)\times \cdots \times U(1)$. In \eqref{quiverkij}, $\xi_i$ and $u_i$ are FI parameters and real mass parameters for chiral multiplets. %We do not give it superpotential.
For early discussions on $\T_{A,N}$ theories see e.g. \cite{Terashima:2014aa}.
It is easy to write down their sphere partition functions
\begin{align}\label{partfunquiver}
Z_{S_b^3}^{\T_{A,N}}  = \int \prod\limits_{i=1}^{N}dx_i \,  e^{ \sum\limits_{i,j=1}^N - i\, \pi  \,  k_{ij} x_i x_j  + 2 \, i\, \pi \, \xi_i x_i  }  \prod\limits_{ i=1}^N s_b\Big( \frac{  i\, Q} {2}  +x_i +  \frac{u_i}{2}   \Big)   \,.
\end{align}
%$\T_{A,N}$ theories are basically $N$ copies of $U(1)-[1]$ coupled together  by mixed Chern-Simons levels $k_{ij}$, and have equal color number and flavor number $N=N_C=N_F$.
Note that if one shifts $x_i$ and defines $\tilde\xi_i$ as follows
\begin{align}
x_i\rightarrow -x_i- \frac{u_i}{2} \,,~~\xi_i = - \tilde\xi_i -\frac{1}{2} \sum\limits_{j=1}^{N}  k_{ij} u_j  \,,
\end{align}
then \eqref{partfunquiver} simplifies to
\begin{align}\label{partfuncT_{AN}}
Z_{S_b^3}^{\T_{A,N  }}  
=\int \prod\limits_{i=1}^{N}dx_i \,  e^{\sum\limits_{i,j=1}^{N}- i\, \pi  \,  k_{ij} x_i x_j  + 2 \, i\, \pi \, \tilde\xi_i x_i  }  \prod\limits_{ i=1}^{N} s_b\Big( \frac{  i\, Q} {2}  -x_i  \Big) 
\end{align}
where real mass parameters $u_i$ are absorbed into shifted FI parameters $\tilde\xi_i$. Therefore we use \eqref{partfuncT_{AN}} as the sphere partition functions of $\T_{A,N}$ theories in the following sections.
Note that if Chern-Simons levels are diagonal $k_{ij}=k _{i} \delta_{ij}$, then $\T_{A,N}$ theories reduce to $N$ copies of independent building blocks
\begin{align}
U(1)_{k_{1}}-[1]~~\oplus~~ U(1)_{k_{2}}-[1] ~~\oplus~ \cdots\oplus~~ U(1)_{k_{N}}-[1]  \,.
\end{align}
In this paper we focus on symmetric CS levels $k_{ij}=k_{ji}$. We find $\T_{A,N}$ theories are very useful for mirror symmetry, and we will show in section 3 that $U(1)_k+N_C \mathbf{C}+ N_{AC} \mathbf{C}$ and some other theories can be transformed into  certain $\T_{A,N}$ theories.

\subsection{Effective superpotential}

After compactifying on a circle $S^1$, 3d $\N=2$ gauge theories can be viewed as 2d $\N=(2,2)$ sigma models with infinitely many KK modes. 
As shown in \cite{Dimofte:2011ju,Dimofte:2011aa,Gadde_2014,Yamazaki:2012aa},  the vortex partition function, sphere partition function, and superconformal index have the same asymptotic expansion in the semiclassical limit $\hbar \rightarrow 0$,
\begin{align}\label{semiclassical}
Z^{\text{vortex}}_{  \mathbb{R}^2 \times S^1  }\,,~ Z_{S_b^3} \,,~ Z_{S^2 \times S^1 }  \sim 
\int \prod\limits_i d\, x_i~
 e^ { \frac{1}{   \hbar   } {  \Weff_{3d \,\mathcal{N}=2 }  ( \mathbf{\xi},\, \mathbf{x})  }+ O(\hbar)  }   \,,
\end{align}
where we have ignored some constant terms. 
The equivariant parameter is related to the quantum parameter $\hbar$ as follows:
\begin{align}
Q = \frac{ \log(q)}{ 2 \pi b \, i }\,, \quad \hbar= 2 \pi \,i\, b^2\,,~~q=e^{\hbar}=e^{ 2\pi \, i \, b^2 } \,.
\end{align}
For $\T_{A,N}$ theories, if we redefine parameters for each gauge node $U(1)_i$,
\begin{align}
x_i =: \frac{ \log \big(-\frac{ y_i}{ \sqrt{q}}  \big)  } {   -2  \pi b}\,,
\end{align}
then the associated twisted effective superpotentials can be obtained by taking the semiclassical limit $\hbar\rightarrow 0$ and using \eqref{sbasym}; this yields
\begin{align}\label{WeffQ}
\Weff_{\T_{A,N} }  ( k_{ij},\mathbf{\xi}, \mathbf{y}  )=\sum\limits_{i=1}^{N_f} \Li_2 (y_i)+\xi^{eff}_{i} \, \log \,y_i+ \sum\limits_{i,j=1}^{N_f} \frac{ k^{eff}_{ ij} }{2} \log\, y_i \, \log\, y_j \,,
\end{align}
where polylogarithm functions $\Li_2(y_i)$ come from contributions of chiral multiplets, $k^{eff}_{ij}$ are effective CS level matrices, and $\xi^{eff}_i$ are effective FI parameters, which are related to bare parameters
\begin{align}\label{effFI}
k^{eff}_{ij}&=k_{ij}+\frac{1}{2} \delta_{ij}   \in  \mathbb{Z}  \,, \\
\xi_i^{eff}&=2 \, \pi b\,\tilde\xi_i+ i\, \pi (1-b \,Q) \sum\limits_{j=1}^{N_f} k_{i j} +\frac{i \pi}{2} \\
&= -2 \pi\, b \,\xi_i + \sum\limits_{j=1}^{N_f}   k_{ij}  \Big(i \pi  -\pi \, b\, u_j - \frac{ \log(q)}{2}  \Big) +\frac{i \pi}{2} \,.
\end{align}
To avoid mistakes, we remind that for symmetric CS terms
\begin{align}
\sum\limits_{i,j}\frac{ k^{eff}_{ ij} }{2} \log\, y_i \, \log\, y_j =\sum\limits_{i} \frac{ k^{eff}_{ ii} }{2} (\log\, y_i)^2 + \sum\limits_{i<j}  k^{eff}_{ ij}  \log\, y_i \, \log\, y_j \,.
\end{align}
Moreover, in \cite{Nekrasov:2009uh, Nekrasov:2009ui} it is shown that the Coulumb branch moduli space $\mathcal{M}_C$ is defined by vacuum equations
\begin{align} \label{vacuaeqn}
\mathcal{M}_C: \quad e^{ y_i  \frac{d \,\Weff}{ d\, y_i}} =1\,, ~ \text{for}~\forall\, i =1, \dots, N \,.
\end{align}
Substituting \eqref{WeffQ} into  \eqref{vacuaeqn} we get
\begin{align}
\mathcal{M}_C: \quad  e^{\, \xi_i^{eff} }  \cdot \prod\limits_{j=1}^N y_j^{k_{ij}^{eff}}  +\,y_i-1=0\,,\quad \forall~ i=1,\dots, N  \,.
\end{align}
%This Coulumb moduli space  $\mathcal{M}_C$ is called classical curve or mirror curve for knot theory. It can be lifted to quantum version, see \cite{Larraguivel:2020sxk} for recent work.
The Hessian matrix of $\Weff$  can also be computed
\begin{align}
	\mathbf{H}\left(\Weff \right)_{i\, j} := \frac{ d^2\, {\Weff} }{ d\, \log\, y_i  \, \log\, y_j}
= \, k_{ij}^{eff} + \delta_{ij} \frac{y_i}{1-y_i} 
\,,\quad \forall~ i\,,j=1,\dots, N  \,.
\end{align}
The vortex partition functions of $\T_{A,N}$ theories can be conjectured by comparing \eqref{WeffQ} with superpotentials in  \cite{Panfil:2018faz}; this implies that they should have the following form
\begin{align}\label{vortexeff}
Z^{\text{vortex}}_{\T_{A,N}} =\sum_{d_1,...,d_N=0}^{\inf}  
(-\sqrt{q})^{ \sum\limits_{i,j=1}^N k^{eff}_{ij}d_i d_j} 
\frac{ 
	x_1^{d_1}\dots x_N^{d_N}
} 
{ 
	(q,q)_{d_1}\dots (q,q)_{d_N} }
\,	
%& x_i :=(-1)^{k^{eff}_{ii}} e^{\xi_i^{eff}} \,. 
\end{align}
where $x_i :=(-1)^{k^{eff}_{ii}} e^{\xi_i^{eff}} $ and q-Pochhammers is defined as $(x;q)_n:= \prod_{i=0}^{n-1}(1-x q^i) $.
One can also factorize sphere partitions to obtain vortex partition  functions using the factorization property found in \cite{Pasquetti:2011fj, Beem:2012mb}. We note that integers $d_i$ have physical meaning.  The poles of the partition function \eqref{partfuncT_{AN}} are located at $x_i=-d_i \,b-h_i/b$, where $d_i$ and $h_i$  are degrees of the North pole and the South pole on a three sphere $S_b^3$,  and are positive integers. In the semi-classical limit $b\rightarrow0$, $h_i$ are restricted to be zero and $d_i$ are positive integers.

\subsection{Open Gopakumar-Vafa formula}

There are intricate relations between prepotentials and superpotentials. In this section we clarify these relations and  discuss formulas encoding open BPS invariants. 

Prepotentials of 3d $\N=2$ gauge theories play a similar role to prepotentials of 5d $\N=1$ gauge theories. The prepotential of a 3d gauge theory on a surface defect $\mathbb{R}^2_{\epsilon_1}\times S^1$  is defined as  \cite{Nekrasov:2002qd,Nekrasov:2009rc,Shadchin_2007,Nakajima_2005} 
\begin{align}\label{prepotential}
\W_{\mathbb{R}^2\times S^1}  =\lim\limits_{\epsilon_1,\epsilon_2 \rightarrow0}  ~\epsilon_2 \, \log\, Z^{\text{vortex}}_{\mathbb{R}^2\times S^1}  \,
\end{align}
where $\epsilon_{1,2} $ are the $\Omega$-deformation parameters. 
In \cite{Nekrasov:2009uh, Nekrasov:2009ui}, the relations between prepotentials and the quantum integrable system have been found.
If we relate $\epsilon_1 $ to Plank constant $\hbar$ by $\hbar=R \, \epsilon_1$, then the combination of \eqref{semiclassical} and \eqref{prepotential} gives rise to
\begin{align}\label{prosuper}
e^{ \frac{  \W_{\mathbb{R}^2\times S^1}}{ \hbar }   } =
\int \prod\limits_{i}^{} d\, x_i~e^ { \frac{1}{   \hbar   } {  \Weff_{3d \,\mathcal{N}=2 }  (  k_{i,j},\, \mathbf{\xi},\, \mathbf{x})  }  }  \,.
\end{align}
%For discussion on prepotentials and quantum integrable system, see .
%where $\hbar=R \, \epsilon_1$. 
%The integral in \eqref{prosuper} only receives contributions from saddle points of vacua equations, which arecalled solutions. In terms of FI parameters these solutions are a branch of discrete points. The integral in \eqref{prosuper} is the summation of these contributions from saddle points if we ignore constant factors from Hessian matrix. For $\T_{A,N}$ theories, we denote saddle points by $\{ y_{o,i}(\xi)\}$ and find
%\begin{align}
%e^{ \frac{  \W_{\mathbb{R}^2\times S^1}}{ \hbar }   } =
%e^ { \sum\limits_{\{y_{o}\} } \frac{ 1  }{  \hbar   } {   \Weff_{3d \, \mathcal{N}=2 }  ( k_{ij}, \mathbf{\xi}, \mathbf{y}) } }  
%= e^{\frac{1}{ \hbar}  \sum\limits_{y_{o,i}}     \Li_2(y_{o,i})   } \cdot
%e^{\frac{1}{\hbar} \big(\sum\limits_{y_{o,i}} \xi^{eff}_i \log \, y_{o,i} +
%	\sum\limits_{y_{o,{i}},y_{o,{j}}}  k^{eff}_{ij} \log \, y_{o,i} \cdot \log \, y_{o,j}
%	\big)  } 	\,.
%\end{align}

Thanks to geometric engineering, the vortex partition functions of 3d $\mathcal{N}=2$ theories can be interpreted as partition functions of open topological strings, which therefore satisfy a refined open Gopakumar-Vafa formula on $\Omega$-background; for more details see \cite{Dimofte:2010tz,Cheng:2021aa}. This formula asserts that the vortex partition functions can be expanded as
\begin{align}\label{openGV}
\begin{split}
Z_{\mathbb{R}^2\times S^1}^{\text{vortex}}  &=  \, \exp \Bigg[
\sum\limits_{\mathcal{C} \in H_2(X,L, \mathbb{Z})} 
\sum\limits_{J,r \in \mathbb{Z}/2} 
\sum\limits_{n=1}^{\inf}
\frac{  (-1)^{2 J+2 r}     q^{   n J   }  \big(   \frac{t}{q}   \big)^{ n \, r}   
	N_{\mathcal{C}}^{(J,r)}
}
{  n \left( q^{ \frac{n}{2} } -   q^{ -\frac{n}{2} }   \right)   }   e^{-n R\, T_{\mathcal{C}}}  
\Bigg]   \\
&=  \, \PE  \Bigg[
\sum\limits_{{\mathcal{C}} \in H_2(X,L, \mathbb{Z})}
\sum\limits_{J,r \in \mathbb{Z}/2} 
\frac{  (-1)^{2 J+2 r}     q^{    J   }  \big(   \frac{t}{q}   \big)^{  \, r}   
	N_{\mathcal{C}}^{(J,r)}}
{     \left( q^{ \frac{1}{2} } -   q^{ -\frac{1}{2} }   \right)     }e^{-R\,T_{\mathcal{C}}}
\Bigg]\,,
\end{split}
\end{align}
where $N_{\mathcal{C}}^{(J,r)}$ are degeneracies of vortex particles and $t=e^{ +R\, \epsilon_1  },\,q =e^{-R\, \epsilon_2}$ parametrize the $\Omega$-background 
\footnote{In the second line of \eqref{openGV},  $\PE[\cdots]$ stands for the plethystic exponential function
\begin{align}
\PE\big[ f( \cdot )\big]:=\text{exp}\Bigg[  \sum\limits_{n=1}^{\inf}
\frac{f(\cdot^n )}{ n}
\Bigg]\,,
\end{align}}.
  The variables $e^{-R\,T_\mathcal{C} }$ are the open K\"ahler parameters for relative 2-cycle ${\mathcal{C}} \in H_2 (X, L, \mathbb{Z})$, and $T_{\mathcal{C}}$ are their volumes, namely the masses of open M2-branes wrapped on ${\mathcal{C}}$, and $R$ is the radius of $S^1$. From the perspective of topological strings, refined open BPS invariants  $N_{\mathcal{C}}^{(J,r)}$ are degeneracies of BPS states (vortex particles) engineered by open M2-branes ending on a M5-brane wrapping a special Lagrangian submanifold $L$ in a Calabi-Yau threefold $X$,  and $(J,r)$ are combinations of charges for the rotation symmetry on $\mathbb{R}^2$ and the $R$-symmetry. %More details on refined open BPS invariants can be found in \cite{Cheng:2021aa}. 
 %for explicit calculations of refined open BPS invariants, in particular $U(1)_k+N_C \mathbf{C}+N_{AC}\mathbf{AC}$ that we will discuss in section \ref{section-3}.

By using the open GV formula, one can find the relations between prepotentials and holomorphic disk potentials. 
Substituting the vortex partition function \eqref{openGV} into \eqref{prepotential} one gets
\begin{align}\label{prepoW}
\W_{\mathbb{R}^2\times S^1}  &=\lim\limits_{\epsilon_1, \epsilon_2 \rightarrow0}  ~\epsilon_2 \, \log\, Z_{\mathbb{R}^2\times S^1}^{\text{vortex}}  \nn\\
&= -\frac{1}{R}  \sum\limits_{{\mathcal{C}} \in H_2(X,L, \mathbb{Z})} 
\sum\limits_{J,r \in \mathbb{Z}/2}  
(-1)^{2 J+2 r}   N_{\mathcal{C}}^{(J,r)}  \Li_2\big(    e^{ -R\, T_{\mathcal{C}} }  \big)  \,.
\end{align}
Expanding the polylogarithm function $\Li_2(z):=\sum\limits_{n=1}^{\inf} \frac{z^n}{n^2}$, the result \eqref{prepoW} takes the form
\begin{align}\label{RWr2s1}
-R \, \W_{\mathbb{R}^2\times S^1}  &=
\sum\limits_{n=1}^{\inf}
\sum\limits_{{\mathcal{C}} \in H_2(X,L, \mathbb{Z})} 
\sum\limits_{J,r \in \mathbb{Z}/2}  
(-1)^{2 J+2 r}   N_{\mathcal{C}}^{(J,r)} \, 
\frac{ e^{-n\,R \, T_{\mathcal{C}}}    }{ n^2 }\,,	
\end{align}
which has the same form as the holomorphic disk potential encoding Ooguri-Vafa invariants in the topological A-model (see \cite{Ooguri:1999bv})
\begin{align}\label{Wopen}
\mathcal{W}_{\text{open}}=
\sum\limits_{{\mathcal{C}} \in H_2(X,L, \mathbb{Z})} 
N^{OV}_{{\mathcal{C}}}  \Li_2\big(    e^{ -R\, T_{\mathcal{C}} }  \big)=\sum\limits_{n=1}^{\inf}
\sum\limits_{{\mathcal{C}} \in H_2(X,L, \mathbb{Z})} 
N^{OV}_{{\mathcal{C}}}\, \frac{ e^{-n\,R \, T_{\mathcal{C}}}    }{ n^2 }   \,.
\end{align}
Therefore, propotentials in 3d $\N=2$ theories are equivalent to holomorphic disk potentials
\begin{align}
-R \, \W_{\mathbb{R}^2\times S^1}  =\mathcal{W}_{\text{open}}  \,,
\end{align}
and classical Ooguri-Vafa invariants can be represented as the summations of refined open BPS invariants\footnote{ Because of this, $N_{\mathcal{C}}^{(J,r)} $ are also called refined Ooguri-Vafa invariants e.g.  in \cite{Cheng:2021aa}.  }
\begin{align}
N^{OV}_{{\mathcal{C}}}=\sum\limits_{J,r \in \mathbb{Z}/2}  
(-1)^{2 J+2 r}   N_{\mathcal{C}}^{(J,r)}   \,.
\end{align}
%for open M2-branes wrapping on relative cycles ${\mathcal{C}} \in H_2 (X, L, \mathbb{Z})$.
Note that the disk potential is classical and can be expressed as an integral in the B-model
\begin{align}
\mathcal{W}_{\text{open}}=\int \log\,y~ \frac{d x}{x}
\end{align}
where $\log\, y$ is the differential one-form on the mirror curve (see \cite{Aganagic:2000gs,Aganagic:2001nx}).
We emphasize that the prepotentials $\W_{\mathbb{R}^2\times S^1}$ are not complete at decompactification limit $R \rightarrow \inf$. Following the treatment in \cite{Hayashi:2019aa}, we define the complete prepotential for 3d $\N=2$ gauge theory in this limit
\begin{align}
\W_{\mathbb{R}^2\times S^1}^{ \text{Complete} }:=\lim\limits_{R\rightarrow +\inf}  \frac{1}{R} \W_{\mathbb{R}^2\times S^1}   \,,
\end{align}
which takes the form
\begin{align}
\W_{\mathbb{R}^2\times S^1}^{ \text{Complete} }  &=
- \frac{1}{ 2}
\sum\limits_{{\mathcal{C}} \in H_2(X,L, \mathbb{Z})} 
\sum\limits_{J,r \in \mathbb{Z}/2}  
(-1)^{2 J+2 r}   N_{\mathcal{C}}^{(J,r)}  \,    \text{\textlbrackdbl}  T_{\mathcal{C}} \text{\textrbrackdbl}   ^2 
\end{align}
where we used  \eqref{limitLi2}.
Furthermore, refined open BPS invariants can be resummed into different invariants in various limits.
In the Nekrasov-Shatashvili (NS) limit $\epsilon\neq0, \epsilon_2 =0$ \cite{Nekrasov:2009ui}, using GV formula \eqref{openGV} we get
\begin{align}
\lim\limits_{\epsilon_2 \rightarrow0}  ~\epsilon_2 \, \log\, Z_{\mathbb{R}^2\times S^1}  = -\frac{1}{R}  \sum\limits_{{\mathcal{C}} \in H_2(X,L, \mathbb{Z})} 
\sum\limits_{J,r \in \mathbb{Z}/2}  
(-1)^{2 J+2 r}   N_{\mathcal{C}}^{(J,r)}  \Li_2\big(  t^r\,  e^{ -R\, T_{\mathcal{C}}}  \big)  \,
\end{align}
which implies that $ N_{\mathcal{C}}^r:=\sum\limits_{J \in \mathbb{Z}/2}  
(-1)^{2 J}   N_{\mathcal{C}}^{(J,r)}  $ are the invariants in the NS limit. 
%It is obvious from \eqref{openGV} that 
In the unrefined limit $\epsilon_1=\epsilon_2$, refined formula \eqref{openGV}  reduces to unrefined formula and we identify $ N_{\mathcal{C}}^J:=\sum\limits_{r \in \mathbb{Z}/2}  
(-1)^{2 r}   N_{\mathcal{C}}^{(J,r)}  $ as the unrefined invariants. Note that $N_{\mathcal{C}}^{(J,r)} $ can only be positive integers, while $N_{\mathcal{C}}^J$ can be either positive or negative integers.

\subsection{Mirror transformation group}

From the perspective of in 3d-3d correspondence,  mirror symmetry corresponds to a change of triangulation of three manifolds that engineer 3d $\N=2$ gauge theories  \cite{Dimofte:2011ju,Terashima:2011qi}.  It	can also be interpreted as a functional Fourier transformation on the partition function \cite{Kapustin:1999ha}, which is  called
mirror transformation in this note. The mirror transformation for 3d $\N=2$ gauge theories with superpotentials was used to derive dualities, e.g., in \cite{Benvenuti:2016wet}. Here we discuss its application to $\T_{A,N}$ theories.  We start from the most basic example, namely the duality between $U(1)_{1/2} + \mathbf{C}$ and a chiral multiplet with Chern-Simons level $-1/2$:
\begin{align}\label{mirrorsymmetry}
(1)_{\frac{1}{2}}-[1] ~~  \xlongleftrightarrow{\text{mirror symmetry}}~~ [1]_{ -\frac{1}{2}  }-[1] \,.
\end{align}
The corresponding partition functions are equivalent
\begin{align}
Z_{S_b^3}^{ (1)_{1/2}-[1] }= Z_{S^3_b}^{[1]_{-1/2}-[1]} \,,
\end{align}
or more explicitly,
\begin{align}\label{oldms}
\int dy\, e^{- \frac{ i\,\pi  }{ 2 } y^2   } e^{ 2 \pi \,i \, \left(\frac{ i\, Q  }{4  }  -z \right)y  } s_b\big(  \frac{ i\, Q}{ 2}-y \big) = e^{ \frac{ i\, \pi }{2}  \left( \frac{ i\, Q}{2} -z\right)^2   } s_b\big(  \frac{ i\, Q }{2} -z  \big)  \,.
\end{align}
This is a mathematical identity presented in \cite{Faddeev_1995,Faddeev2001}, which implies that any double-sine function $s_b(\dots)$ can be replaced by a contour integral. This is analogous to gauging $U(1)$ flavor symmetry 
\begin{align}\label{mirror1}
[1]-[1]  \xrightarrow{\text{mirror transformation}} (1)-[1]\,.
\end{align}
In terms of sphere partition functions, this replacement takes form
\begin{align}\label{mirroridentity}
s_b\big(  \frac{ i\, Q }{2} -z  \big)  ~ \xrightarrow{\text{mirror transf.}}~e^{ -\frac{ i\, \pi }{2}  \left( \frac{ i\, Q}{2} -z\right)^2   }      \int dy\, e^{- \frac{ i\,\pi  }{ 2 } y^2   } e^{ 2 \pi \,i \, \left(\frac{ i\, Q  }{4  }  -z \right)  } s_b\big(  \frac{ i\, Q}{ 2}-y \big)  \,.
\end{align}
Note that the double since functions, as one-loop contributions of chiral multiplets, can be regarded as basic units for mirror transformations.

Moreover, mirror symmetry turns out to be $ST$ operation from the $SL(2,\mathbb{Z})$ viewpoint,  when acting on the Lagrangian of 3d Chern-Simons theory, as found by Witten in \cite{Witten:2003ya},  so one can also use $ST$ to stand for mirror symmetry. 
%Nothing stops us from applying mirror symmetry on chirals more than one time. 
After performing mirror symmetry on the quiver $(1)_k-[1]$ only once, we get a new quiver $(1)'_{k'}-[1]$ :
\begin{align}
ST: \quad(1)_k-[1]  ~~ \xrightarrow{ST} ~~ (1)-\big(  \, (1)'-[1] \, \big)    ~~\xrightarrow{\text{integrate out } (1)_k } ~~  (1)'_{k'}-[1]  \,,
\end{align}
where the original gauge group $(1)_k$ was integrated out to get the new quiver with CS level $k'$ and new FI parameters $\xi'$. This transformation does not change partition functions
$
Z^{ (1)_k-[1] }_{S_b^3} =Z^{ (1)'_{k'}-[1] }_{S_b^3} 
$.
After performing mirror symmetry twice we get another quiver $(1)''_{k''}-[1]$  :
\begin{align}
(ST)^2: \quad &(1)_k-[1]  ~~ \xrightarrow{ST} ~~ (1)-\big(  \, (1)'-[1] \, \big) ~~ \xrightarrow{ST} ~~  (1)-\big(  \, (1)'- \big( (1)''- [1]    \big) \, \big)    \nn \\
&\xrightarrow{\text{integrate out~}  {(1)}, ~ {(1)}'}  ~~  (1)''_{k''}-[1]  \,.
\end{align}
The corresponding partition functions are also equal 
$
Z^{ (1)_k-[1] }_{S_b^3} =Z^{ (1)''_{k''}-[1] }_{S_b^3} 
$.
Furthermore, after performing mirror transformation for the third time, we return to the original theory
% with the same CS levels $k_{ij}$ and FI parameters $\xi_i$ :
\begin{align}
(ST)^3: \quad (1)_k-[1]  ~~ \xrightarrow{ST} ~~ (1)'_{k'}-[1]  ~~ \xrightarrow{ST}  (1)''_{k''}-[1] ~~ \xrightarrow{ST} ~~(1)_{k}-[1]  \,,
\end{align}
in agreement with the relation
$
(ST)^3=1 
$.

Analogously  we can perform mirror transformations on each building block of $\T_{A,N}$ theories, as illustrated by the following example
\begin{align}
&
\left(
\begin{array}{l}
(1)-[1]\\
(1)-[1]\\
\quad \dots\\
(1)-[1]
\end{array}  \right)_{k_{ij}, ~\xi_i}
\xrightarrow{(ST, ST,\dots, 0)}
\left(
\begin{array}{l}
(1)-((1)'-[1])\\
(1)-((1)'-[1])\\
\quad \dots\\
(1)-[1]
\end{array} 
\right)
\xrightarrow{\text{integrate out } (1)  } 
&
\left(
\begin{array}{l}
(1)'-[1]\\
(1)'-[1]\\
\quad \dots\\
(1)-[1]
\end{array}
\right)_{k'_{ij}, ~\xi'_i}  \,,
\end{align}
where we perform mirror transformations on some gauge nodes of $U(1)\times U(1)\times \dots \times U(1)$. After integrating out old gauge parameters, we get another $\T'_{A,N}$ theory with CS levels $k'_{i,j}$ and FI parameters $ \xi'_i$. 
We find that mirror transformations, acting on various $U(1)_i$ gauge nodes, commute with each other, which implies the following
%Hence mirror transformations on gauge node $U(1)_i$ 
equivalence relation 
\begin{align}\label{equivalence1}
\mathbf{(n_1, n_2, \dots n_i,\dots, n_N ) }\sim \mathbf{(n_1, n_2, \dots n_i+3,\dots, n_N ) },~~\forall\, i =1, \dots, N
\end{align}
where we introduce a shorthand notation
\begin{align}
\mathbf{(n_1, n_2, \dots n_i,\dots n_N )} := \big(  \,(ST)^{n_1},(ST)^{n_2},\dots, (ST)^{n_N}     \big)  \,.
\end{align}
Since $k_{i,j}$ is symmetric, one can exchange its rows and columns
\begin{align}
%\text{If} \quad x_i \leftrightarrow x_j  \,,~~\text{then}~
 k_{i,l} \leftrightarrow k_{j,l} \,,~k_{l,i} \leftrightarrow k_{l,j} ,~\text{for}~\forall ~l =1,\dots, N \,,
\end{align}
by exchanging parameters $x_i \leftrightarrow x_j $  for gauge nodes $U(1)_i$ and $U(1)_j$.
%This exchange equivalence can be represented as exchange symmetry between mirror transformations
This gives another equivalence relation 
%of mirror transformations
\begin{align}\label{equivalence2}
\mathbf{n_i}  \leftrightarrow \mathbf{n_j}  \,.
\end{align}
Composing equivalence relations \eqref{equivalence1} and \eqref{equivalence2}, we introduce a group of mirror transformations 
\begin{align}\label{defmirrorsymgroup}
\mathcal{H} (\T_{A, N})& :=\{ \mathbf{(n_1,n_2,\dots, n_N) } ~|~\mathbf{n_i} \in \{0,1,2\}, ~ \mathbf{n_i}   \geq \mathbf{n_j} ~\text{if}~ i \leq j \,,  ~\forall~ i, j =1 ,2 ,\dots, N   \}   \nn \\ 
&~ = \{ \mathbf{(0,0,\dots,0),~ (1,0,\dots,0), \dots, (2,2,\dots,2)  }  \} 
\end{align}
with a finite number of elements 
$
\frac{ N^2 +3 N +2 }{2} 
$.
This group is additive under mirror transformations 
\begin{align} \label{pertn1n2nN}
&\mathbf{(i_1,i_2 ,\dots, i_N)}:~  \mathbf{(n_1, n_2, \dots n_i,\dots n_N ) } \rightarrow 
\mathbf{(n_1+i_1, n_2+i_2,\dots, n_N+i_N ) }\,,
\end{align}
which implies that $\mathcal{H}(\T_{A, N})$ has a group structure with addition defined as
\begin{align}
(\mathbf{i_1,i_2 ,\dots, i_N}   ) + \mathbf{(n_1, n_2, \dots n_i,\dots n_N ) } =
\mathbf{(n_1+i_1, n_2+i_2,\dots, n_N+i_N ) } \,.
\end{align}
%Since $\mathcal{H}(\T_{A, N})$ is a finite group, 
Note that each element $(\mathbf{i_1,i_2 ,\dots, i_N}   ) $ can be regarded as a permutation on the group $\mathcal{H}(\T_{A, N})$.
Although mirror  transformations produce many mirror dual theories with different Chern-Simons levels and FI parameters, their partition functions are equal up to some irrelevant factors
\begin{align}
Z^{ \T_{A,N}}_{S_b^3} =Z^{ \T_{A,N}}_{S_b^3} // \mathcal{H} (\T_{A,N}) \,.
\end{align}
%and hence these theories are equivalent.
Note that mirror transformations may give rise to effective mixed CS levels $k_{i,j}^{eff}$ with fractional (non-integer) numbers; in this case, the associated theories should be regarded as inconsistent and ignored, as not meeting the parity anomaly constraint $k_{i,j}^{eff} \in \mathbb{Z}$  \cite{Aharony:1997bx,Intriligator:2013lca}.

Let us denote the original  theory  by $\T[\mathbf{(0, \dots, 0)}]$. %with superpotential  $ \widetilde{\mathcal{W}}^{eff,~ \mathbf{( 0,0,\dots, 0 )} } $. 
Mirror transformation $\mathbf{(i_1,\dots, i_N)}$ acting on it leads to a mirror dual theory $\T[\mathbf{(i_1,\dots, i_N)}]  $ with superpotential $ \widetilde{\mathcal{W}}^{eff,~ \mathbf{(i_1,\dots, i_N)}   } $.
This is therefore a correspondence
\begin{align}\label{1to1}
\mathbf{(i_1,\dots, i_N)}  \xleftrightarrow{\text{one to one}}  \T[ \mathbf{(i_1,\dots, i_N)} ]    \,.
\end{align}
Furthermore, based on \eqref{pertn1n2nN},
$\mathbf{(i_1,\dots, i_N)}$ gives rise to a map between $  \T[\mathbf{(n_1,\dots, n_N)}]$ and $  \T[\mathbf{(n_1+i_1,\dots, n_N+i_N)}]$ for $\forall ~ \mathbf{(n_1,\dots, n_N)} \in \H(\T_{A,N})$:
\begin{align}\label{permu}
%(  d_1^{i_1},d_2^{i_2},\dots, d_N^{i_N}):~~ 
&\mathbf{(i_1,\dots, i_N)}:~
\T[\mathbf{(n_1,\dots, n_N)}] 
\rightarrow
\T[\mathbf{(n_1+i_1,\dots, n_N+i_N)}]\,, 
\end{align}
which can be viewed as the mirror map between mirror dual theories, describing the relations between effective CS levels and FI parameters for dual theories.
Since a group of mirror transformations is finite, each $ \mathbf{(i_1,\dots, i_N)}$ can be regarded as a permutation. We can think of
%mirror maps do not depend on specific theory $\T[  \mathbf{(n_1,\dots, n_N)}]$ and
 any mirror dual theory $\T[  \mathbf{(n_1,\dots, n_N)}]$ as the original theory, and act on it with all mirror transformations in $\mathcal{H}( \T_{A,N})$ to obtain a chain of mirror dual theories. In addition, as we mentioned before, the parity anomaly imposes constraints
$
k^{eff}  \in \mathbb{Z}$,
hence only a subset of mirror dual theories are consistent, and we denote them by
\begin{align}
\textbf{Class}(\T_{A,N}  )  := \{
\T[\mathbf{(n_1+i_1,\dots, n_N+i_N)}]  ~\text{with $k^{eff}_{ij} \in \mathbb{Z}$}  ~\,,
~
\forall ~ \mathbf{(i_1,i_2,\dots,i_N)} \in \mathcal{H} (\T_{A,N}) \,
 \}   \,.
\end{align}
We summarize that for any $ \mathbf{(i_1,\dots, i_N)} \in \H(\T_{A,N})$, there are correspondences as follows:
\begin{align}\label{corres}
\mathbf{(i_1,\dots, i_N)}  \xleftrightarrow{\text{one to one}}  \T[\mathbf{(i_1,\dots, i_N)}] 
\xleftrightarrow{\text{one to one}}  \text{permutations}  \xleftrightarrow{\text{one to one}}  \text{mirror maps}   \,,
\end{align}
and mirror dual theory $ \T[\mathbf{(n_1,\dots, n_N)}] $ can be labeled by effective CS levels and FI parameters encoded in effective superpotentials	
\begin{align}
\T[\mathbf{(n_1,\dots, n_N)}] ~ :~\Big(  k^{eff,~\mathbf{( n_1,\dots, n_N )} }_{ij},~  \xi^{eff,~\mathbf{( n_1,\dots, n_N )} }_i  \Big)  \,.   
\end{align}
We will illustrate these relations in examples discussed in section \ref{section-3}.

\subsubsection*{Example}
Consider mirror transformations for the theory
$
\T_{A,2}:~(U(1)-[1])^{ \otimes  2}_{k_{ij}} \,,
$
whose sphere partition function is
\begin{align}\label{originalTa2}
Z^{\T_{A,2}}_{S_b^3}= \int d x_1 dx_2 \, e^{  2 \pi \, i\, ( \tilde\xi_1 x_1+ \tilde\xi_2 x_2)  - i\, \pi (k_{1,1} x_1^2   + 2 k_{1,2} x_1 x_2  + k_{2,2} x_2^2   )  } s_b( 
\frac{i Q}{2}-x_1) s_b(\frac{i Q}{2}-x_2)  \,.
\end{align}
According to \eqref{effFI}, $\T_{A,2}$ theory has the following effective CS levels and FI parameters 
\begin{align}
&k^{eff}_{i,j} =k_{ij}+\frac{1}{2} \delta_{ij} \,, \quad i,j=1,\,2  \,,\\
&\xi_i^{eff}=2 \, \pi b\,\tilde\xi_i+ i\, \pi (1-b \,Q) \sum\limits_{j=1}^{2} k_{i j} +\frac{i \pi}{2}   \,.
\end{align}
We think of  \eqref{originalTa2} as the partition function for the original theory $\T[\mathbf{(0,0)}]$.
Following \eqref{defmirrorsymgroup}, we write its mirror transformation group as
\begin{align}
\mathcal{H}(\T_{A,2})= \{ \mathbf{ (0,0), (1,0),(1,1),(2,0), (2,1),(2,2)}  \}\,,
\end{align}
which corresponds to mirror dual theories $ \{   ~ \T[  \mathbf{0,0}],  \T[  \mathbf{1,0}], \T[  \mathbf{2,0}], \T[  \mathbf{1,1}], \T[  \mathbf{2,1}], \T[  \mathbf{2,2}]~ \}$.
Following mirror maps between dual theories \eqref{permu},  we note that these theories are related by basic mirror transformations $\bf{(1,0)}$ and $\bf{(0,1)}$, as shown in the following commutative diagram
\begin{align}\label{commudiagram}
\includegraphics[width=2.5in]{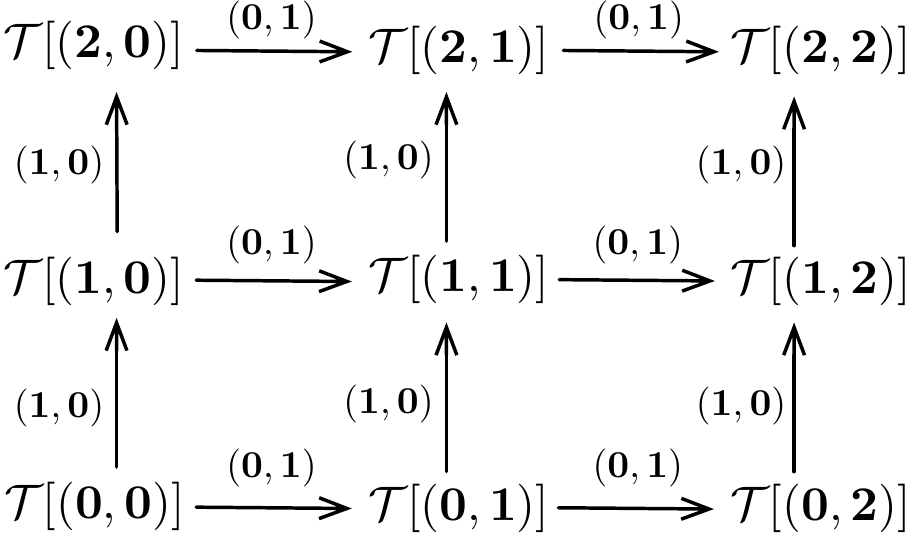} \,.
\end{align}
Each mirror dual theory has associated effective twisted superpotential $\widetilde{\mathcal{W}}^{eff,~ \mathbf{( n_1,n_2)}} $.  
The effective  CS levels $k_{ij}^{eff}$ for all theories in the above diagram  \eqref{commudiagram} read
\begin{align}\label{dataT22}
\begin{split}
	&\T [\mathbf{(0,0) }  ]  ~:~~
\left(
\begin{array}{cc}
	k_{1,1}+\frac{1}{2} & k_{1,2} \\
	k_{1,2} & k_{2,2}+\frac{1}{2} \\
\end{array}
\right)~~	\,, \\
	&   \T  [ \mathbf{(1,0) }]   ~:~~
	\left(
	\begin{array}{cc}
	\frac{2 k_{1,1}-1}{2 k_{1,1}+1} & -\frac{2 k_{1,2}}{2 k_{1,1}+1} \\
	-\frac{2 k_{1,2}}{2 k_{1,1}+1} & \frac{-4 k_{1,2}^2+2 k_{2,2}+k_{1,1} \left(4 k_{2,2}+2\right)+1}{4 k_{1,1}+2} \\
	\end{array}
	\right) \,,\\
		&\T[\mathbf{(0,1) }   ~:~~
		\left(
		\begin{array}{cc}
		\frac{-4 k_{1,2}^2+2 k_{2,2}+k_{1,1} \left(4 k_{2,2}+2\right)+1}{4 k_{2,2}+2} & -\frac{2 k_{1,2}}{2 k_{2,2}+1} \\
		-\frac{2 k_{1,2}}{2 k_{2,2}+1} & \frac{2 k_{2,2}-1}{2 k_{2,2}+1} \\
		\end{array}
		\right)~~\,, \\
			&\T[\mathbf{(2,0) }   ~:~~
			\left(
			\begin{array}{cc}
			\frac{2}{1-2 k_{1,1}} & \frac{2 k_{1,2}}{2 k_{1,1}-1} \\
			\frac{2 k_{1,2}}{2 k_{1,1}-1} & \frac{-4 k_{1,2}^2-2 k_{2,2}+k_{1,1} \left(4 k_{2,2}+2\right)-1}{4 k_{1,1}-2} \\
			\end{array}
			\right) \,,\\
				&   \T[\mathbf{(0,2) }   ~:~~
				\left(
				\begin{array}{cc}
				\frac{-4 k_{1,2}^2+2 k_{2,2}+k_{1,1} \left(4 k_{2,2}-2\right)-1}{4 k_{2,2}-2} & \frac{2 k_{1,2}}{2 k_{2,2}-1} \\
				\frac{2 k_{1,2}}{2 k_{2,2}-1} & \frac{2}{1-2 k_{2,2}} \\
				\end{array}
				\right) \,,\\
					&  \T[\mathbf{(1,1) }   ~:~~
					\left(
					\begin{array}{cc}
					\frac{2 \left(-4 k_{1,2}^2-2 k_{2,2}+k_{1,1} \left(4 k_{2,2}+2\right)-1\right)}{-8 k_{1,2}^2+4 k_{2,2}+k_{1,1} \left(8
						k_{2,2}+4\right)+2} & \frac{4 k_{1,2}}{-4 k_{1,2}^2+2 k_{2,2}+k_{1,1} \left(4 k_{2,2}+2\right)+1} \\
					\frac{4 k_{1,2}}{-4 k_{1,2}^2+2 k_{2,2}+k_{1,1} \left(4 k_{2,2}+2\right)+1} & \frac{2 \left(-4 k_{1,2}^2+2
						k_{2,2}+k_{1,1} \left(4 k_{2,2}-2\right)-1\right)}{-8 k_{1,2}^2+4 k_{2,2}+k_{1,1} \left(8 k_{2,2}+4\right)+2} \\
					\end{array}
					\right) \,,\\
						&  \T[\mathbf{(2,1) }   ~:~~
						\left(
						\begin{array}{cc}
						\frac{2 \left(2 k_{2,2}+1\right)}{4 k_{1,2}^2+2 k_{2,2}-2 k_{1,1} \left(2 k_{2,2}+1\right)+1} & \frac{4 k_{1,2}}{4
							k_{1,2}^2+2 k_{2,2}-2 k_{1,1} \left(2 k_{2,2}+1\right)+1} \\
						\frac{4 k_{1,2}}{4 k_{1,2}^2+2 k_{2,2}-2 k_{1,1} \left(2 k_{2,2}+1\right)+1} & \frac{4 k_{1,2}^2+k_{1,1} \left(2-4
							k_{2,2}\right)+2 k_{2,2}-1}{4 k_{1,2}^2+2 k_{2,2}-2 k_{1,1} \left(2 k_{2,2}+1\right)+1} \\
						\end{array}
						\right)\,,\\
							&  \T[\mathbf{(1,2) }   ~:~~
							\left(
							\begin{array}{cc}
							\frac{4 k_{1,2}^2+k_{1,1} \left(2-4 k_{2,2}\right)+2 k_{2,2}-1}{4 k_{1,2}^2+k_{1,1} \left(2-4 k_{2,2}\right)-2
								k_{2,2}+1} & \frac{4 k_{1,2}}{4 k_{1,2}^2+k_{1,1} \left(2-4 k_{2,2}\right)-2 k_{2,2}+1} \\
							\frac{4 k_{1,2}}{4 k_{1,2}^2+k_{1,1} \left(2-4 k_{2,2}\right)-2 k_{2,2}+1} & \frac{2 \left(2 k_{1,1}+1\right)}{4
								k_{1,2}^2+k_{1,1} \left(2-4 k_{2,2}\right)-2 k_{2,2}+1} \\
							\end{array}
							\right)\,,\\
								&  \T[\mathbf{(2, 2) }   ~:~~
								\left(
								\begin{array}{cc}
								\frac{2-4 k_{2,2}}{-4 k_{1,2}^2-2 k_{2,2}+k_{1,1} \left(4 k_{2,2}-2\right)+1} & \frac{4 k_{1,2}}{-4 k_{1,2}^2-2
									k_{2,2}+k_{1,1} \left(4 k_{2,2}-2\right)+1} \\
								\frac{4 k_{1,2}}{-4 k_{1,2}^2-2 k_{2,2}+k_{1,1} \left(4 k_{2,2}-2\right)+1} & \frac{2-4 k_{1,1}}{-4 k_{1,2}^2-2
									k_{2,2}+k_{1,1} \left(4 k_{2,2}-2\right)+1} \\
								\end{array}
								\right) \,.
							%	\label{dataT22end}
								\end{split}
\end{align}
It is obvious that the equivalence relations	 \eqref{equivalence1} and \eqref{equivalence2} are satisfied, and 
parity anomaly strongly constrains the possible values of $k_{ij}$ in \eqref{dataT22}.

\subsection{Quiver reduction}

Given some specific values of bare CS levels $k_{ij}$,  the effective CS levels $k^{eff}_{ij} $ may be problematic in some cases: the effective CS levels have poles or vanishing determinant
\begin{align}\label{quiverred}
k^{eff}_{ij} =
\begin{pmatrix}
* & * & \dots &*\\
* & \frac{a}{0} & \dots &\frac{c}{0}\\
*& \frac{b}{0}& \dots &\frac{d}{0}
\end{pmatrix}
\quad \text{or} \quad
\text{det}\, k_{ij}=0 \,.
\end{align}
  We call this phenomenon quiver reduction.  For instance, there are quiver reductions for effective CS levels in \eqref{dataT22}, when $k_{1,1}=\pm1/2$, $k_{2_,2}=\pm1/2$, etc.
%Quiver reduction implies some gauge nodes are redundant, which can be integrated out by hand.
Using formula \eqref{integralAJ}, one can see that for the CS levels in \eqref{quiverred},
the contour integral over gauge nodes is not Gaussian, but takes the form of the Dirac delta function that reduces the dimension of the full integral. Namely, quiver reductions imply some gauge nodes are redundant and can be integrated out. 
%In the following sections we will illustrate by more examples. 
%We do not have a clear understanding on quiver reduction, hence leave it for future work. 

\subsection{CS level decomposition and charge vectors}
We can generalize the story to generic $\T_{A,N}$ theories with chiral multiplets of other charges except $\pm 1$. It turns out that charge vectors and CS level matrices for these theories are exchangeable.

Let us start with generic theories with  gauge groups
$
U(1)_1\times U(1) _2\times \dots \times U(1)_N
$
and $N$ chiral multiplets in arbitrary representations. These theories have partition functions of the form
\begin{align}\label{dep1}
Z_{S_b^3} ( \mathbf{K},  \mathbf{P} )
=\int d\, \textbf{x} \,  e^{- i\, \pi  \, \textbf{x}^{{T}} \textbf{K } \textbf{x}  + 2 \, i\, \pi \,   \mathbf{\tilde{\xi}}^T \textbf{x}  }  \prod\limits_{ i=1}^N s_b\Big( \frac{  i\, Q} {2}  -\textbf{P}^T_i \cdot \textbf{x}  \Big)
\,,
\end{align}
where $ \textbf{x}=(x_1; x_2; \dots; x_N )$ is a $N \times1 $ matrix, and $\textbf{P}^T_i$ are charge vectors for chiral multiples.
We define
\begin{align}
\textbf{P}:=(\textbf{p}_1 , \textbf{p}_2,\dots ,\textbf{p}_N  )\,,
\end{align}
where $\mathbf{P}_i=\mathbf{p}_i$ and
$
\textbf{y}:=\textbf{P}^T\,\textbf{x}\,.
$
After this variable transformation, and ignoring the Jacobian matrix, charge vectors can be absorbed into new mixed CS levels and FI parameters, and \eqref{dep1} becomes
\begin{align}\label{dep2}
Z_{S_b^3} ( \mathbf{K}',  \mathbf{1})
&=\int d\, \textbf{y} \,  e^{- i\, \pi  \, \textbf{y}^{{T}} \textbf{K}' \textbf{y}  + 2 \, i\, \pi \,   \mathbf{\tilde{\xi}}'^T \textbf{y}  }  \prod\limits_{ i=1}^N s_b\Big( \frac{  i\, Q} {2}  -y_i  \Big) \,, \\
\textbf{K}'&= (\textbf{P}^{-1}) \cdot  \textbf{K} \cdot (\textbf{P}^{-1})^T
\,,\\
\mathbf{\tilde{\xi}}'& = (\textbf{P}^{-1}) \cdot \mathbf{\tilde{\xi}}  \,.
\end{align}
 If $\textbf{K}$ is symmetric, then $\textbf{K}'$ is also symmetric. 
Both $\textbf{K}$ and  $\textbf{K}'$ can be decomposed in the orthogonal basis and have the same eigenvalues $\mathbf{\Lambda}$
\begin{align}\label{trick1}
\textbf{K}&= \mathbf{Q}^{-1} \mathbf\Lambda \,  (\mathbf{Q}^{-1})^T = \mathbf{Q}^{T} \mathbf\Lambda \,  \mathbf{Q}   \,, \\
\textbf{K}' &=  (\textbf{P}^{-1}) \cdot  \textbf{K} \cdot (\textbf{P}^{-1})^T
= \mathbf{Q'}^{-1} \mathbf\Lambda \,  (\mathbf{Q'}^{-1})^T       \,,    \\
\textbf{Q}'&=\mathbf{Q\, P}  \,.
\end{align}
The partition function \eqref{dep2} is exactly the sphere partition function for $\T_{A,N}$ theory. Therefore, we can turn generic Abelian theories \eqref{dep1} into $\T_{A,N}$ type \eqref{dep2}.
Moreover, with the help of \eqref{trick1}, the form  \eqref{dep1} can also be transformed into theories with diagonal CS levels but complicated charge vectors
\begin{align}\label{dep3}
Z_{S_b^3}  ( \mathbf{\Lambda},   \mathbf{Q\, P }  )
&=\int d\, \textbf{z} \,  e^{- i\, \pi  \, \textbf{z}^{\textbf{T}} \mathbf{\Lambda} \, \textbf{z}  + 2 \, i\, \pi \,   \big(\mathbf{ Q\, \tilde{\xi}} \big)^T\textbf{z}  }  \prod\limits_{ i=1}^N s_b\Big( \frac{  i\, Q} {2}  -  ( \mathbf{Q\, P } )_i^T\cdot  \textbf{z}  \Big) \,, 
\end{align}
where 
$\mathbf{x}=\mathbf{Q}^T \, \mathbf{z}  $. We call it charge vector form.

Based on the above discussion, 	one can transform generic theories \eqref{dep1} into either  $\T_{A,N}$ type theories \eqref{dep2} with mixed CS level $ \mathbf{K}'$ and simple charge vectors $\mathbf{1}$ or charge vector form \eqref{dep3} with diagonal CS level $ \mathbf{\Lambda}$ and complicated charge vectors $\mathbf{Q\, P } $
\begin{align}
%(\text{CS levels}\,,\, \text{ charge vectors}):~~
( \mathbf{K},  \mathbf{P}) \rightarrow ( \mathbf{K}',  \mathbf{1})  ~\text{or}~( \mathbf{\Lambda},   \mathbf{Q\, P }  ) \,.
\end{align}
The associated effective superpotentials for these three forms \eqref{dep1}, \eqref{dep2}, and \eqref{dep3} are equivalent. Hence these three forms of partition functions are supposed to correspond to the same mirror theory class $\textbf{Class}(\T)$.
In this note we only consider the form \eqref{dep2} and leave the charge vector form \eqref{dep3} for future work. 
In addition, if $\mathbf{K}$ is real positive  definite, then it has Cholesky decomposition
$
\mathbf{K}=\mathbf{L}^T \mathbf{L}
$,  and \eqref{dep1} can be turned into another form
\begin{align}
Z_{S_b^3} ( \mathbf{1},  (\mathbf{L}^{-1})^T\mathbf{P} )
=\int d\, \mathbf{x'} \,  e^{- i\, \pi  \, \mathbf{x'}^{{T}}\mathbf{x'}  + 2 \, i\, \pi \,   \left(\mathbf{ (\mathbf{L}^{-1})^T\tilde{\xi}}\right)^T \mathbf{x'}  }  \prod\limits_{ i=1}^N s_b\Big( \frac{  i\, Q} {2}  -\big((\mathbf{L}^{-1})^T\mathbf{P} \big)^T_i \cdot \mathbf{x'}  \Big)
\,,
\end{align}
where $\mathbf{x'}=\textbf{L} \, \mathbf{x}$.

\section{$U(1)_k+ N_C \,\mathbf{C} + N_{AC} \, \mathbf{AC}$} \label{section-3}

\subsection{Brane webs}

\begin{figure}
	\centering
	\includegraphics[width=3in]{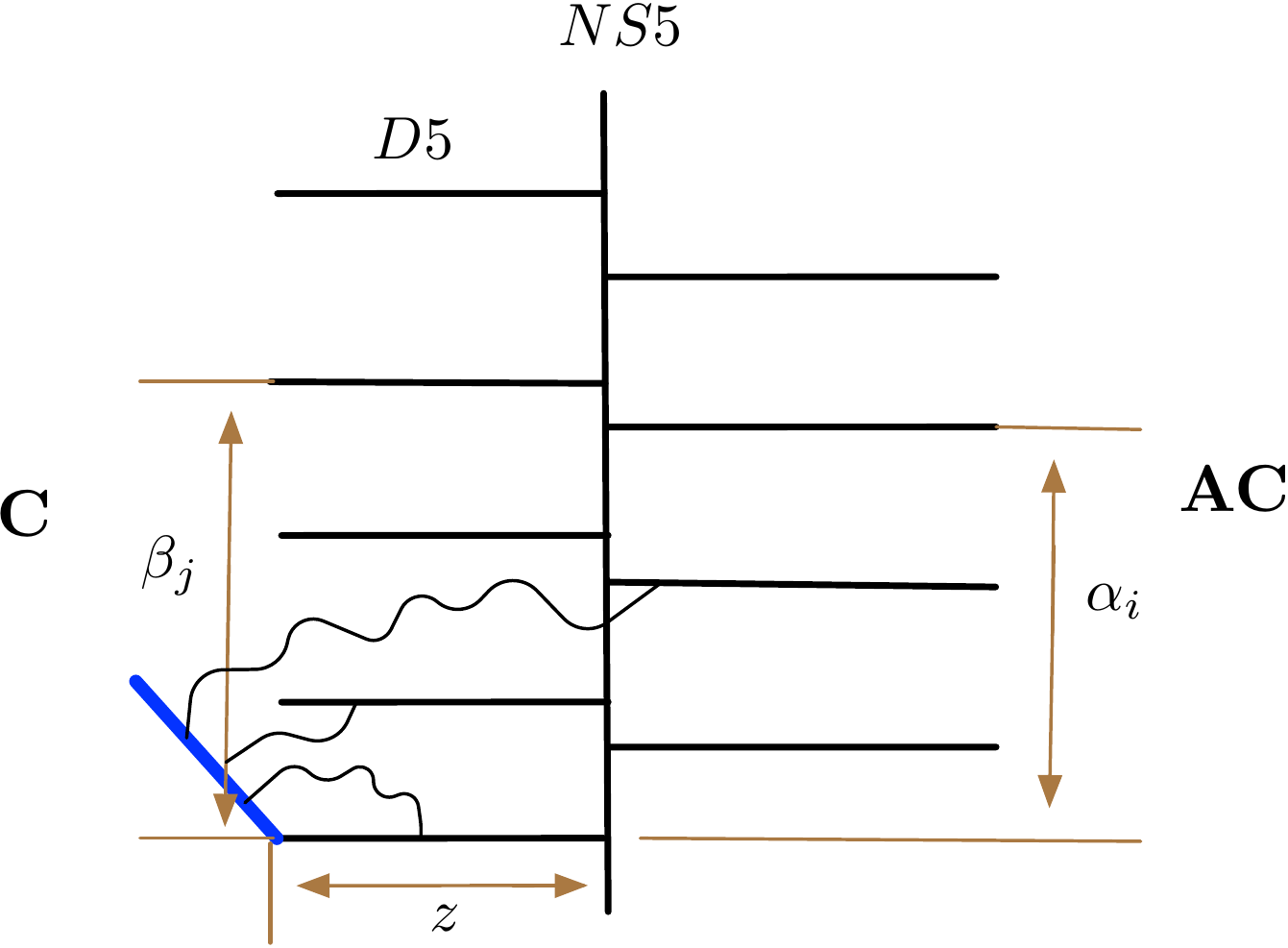} 
	\caption{This diagram is the IIB brane construction for theories $U(1)_k+ N_C \mathbf{C}+ N_{AC} \mathbf{AC}$. The blue line stands for D3-brane as a surface defect. The horizontal lines denote D5-branes, and the vertical line denotes the NS5-brane. The wavy lines denote open strings between the D3-brane and D5-branes. This IIB brane web is dual to toric Calabi-Yau threefold with a Lagrangian brane	through IIB/M-theory duality. 
	}
	\label{fig:strip}
\end{figure}

We denote by $U(1)_k+ N_C \,\mathbf{C} + N_{AC} \, \mathbf{AC}$ the theories that contain gauge group $U(1)$ and $N_C$ chiral multiplets of  charge $+1$ and $N_{AC}$ chiral multiples of charge $-1$. These theories can be engineered as surface defect theories by Higgsing 5d $\N=1$ brane webs. See, e.g., \cite{Kozcaz:2010af,Dimofte:2010tz} for more details.  The corresponding brane configuration in type IIB strings is shown in figure \ref{fig:strip}. 
In this brane web, open strings connecting the D3-brane and D5-brane on the left-hand side of the NS5-brane give rise to fundamental chiral multiplets denoted by $\mathbf{C}$, and the open strings connecting D3-brane and D5-branes on the right-hand side of the NS5-brane give rise to antifundamental chiral multiplets denoted by $\mathbf{AC}$. 
Note that in this brane construction, there is the freedom of putting D3-brane on any D5-branes on the left-hand side of the NS5-brane, which gives rise to the same 3d $\N=2$ theories. 
However, if moving the D3-brane to D5-branes on the right-hand side of the NS5-brane, then $\mathbf{C}$ and $\mathbf{AC}$ are switched; hence, the matter content of the theory is changed,  
so this movement leads to different theories. 
In addition, the string located at the D3--D5-brane intersection is of length zero,and hence the corresponding chiral multiplet is massless \cite{Dimofte:2010tz}. 

 The duality between type IIB strings and  M-theory can be represented in terms of a brane construction and geometric engineering.  From this viewpoint, brane webs correspond to strip Calabi-Yau threefolds, and the associated vortex partition functions are interpreted as open topological string partition functions.
See \cite{Cheng:2018ab,Cheng:2021aa} for discussions on open topological string amplitudes, Higgsing, and Hanany-Witten transitions for $U(1)_k+ N_C \,\mathbf{C} + N_{AC} \, \mathbf{AC}$ theories. 
	
 The $U(1)_k+N_C\mathbf{C}+N_{AC} \mathbf{AC}$ can be rewritten as $\T_{A,N}$ theories by doing mirror transformation $\mathbf{(1,1, \dots, 1)}$ and integrating out the original gauge node $U(1)_k$
\begin{align}\label{equalence}
U(1)_k+N_C\mathbf{C}+N_{AC} \mathbf{AC}  \xlongrightarrow{\mathbf{(1,1, \dots, 1)}} \T_{A,\, N_C+N_{AC}}   \,. %~~\text{with}~N=N_C+N_{AC}
\end{align}
This implies that performing mirror transformation on  $U(1)_k+N_C\mathbf{C}+N_{AC} \mathbf{AC}$ is equivalent to performing mirror transformations on $\T_{A,N_C+N_{AC}}$ theories. We take $U(1)_k+N\mathbf{C}$ theory as an example, 
whose sphere
partition functions can be transformed into $\T_{A,N}$ theories
\begin{align}\label{U1NtoTAN}
Z_{S_b^3}^{   \,(1)_k-[N]\,}  \xlongrightarrow{\mathbf{(1,1, \dots, 1)}}
Z_{S_b^3}^{  \T_{A,N}}\,.
\end{align}
More explicitly, by \eqref{shperepartNC}, the associated sphere partition functions for $U(1)_k+N\mathbf{C}$ take the following form:
\begin{align}\label{sphere1N}
Z_{S_b^3}^{   \,(1)_k-[N]\,}  = \int dx \,  e^{ - i\, \pi  \,  k x^2  + 2 \, i\, \pi \xi x  }  \prod\limits_{ i=1}^N s_b\Big( \frac{  i\, Q} {2}  +x + \frac{u_i}{2}  \Big)  \,,
\end{align}
 which in the semiclassical limit
\eqref{sphere1N} gives the effective superpotential
\begin{align}
\Weff_{{(1)_k-[N]} }  &=\sum\limits_{i=1}^{N} \Li_2 (X Y_i)+\xi^{eff} \, \log \,X+ \frac{ k^{eff} }{2}  \big( \log\, X \big)^2   \,, \\
k^{eff}&= k+\frac{N}{2}\,,~~
\xi^{eff} =\frac{1}{2} \big( i\, \pi N -4b\pi \xi +\log\, \prod\limits_{i=1}^N Y_i  \big)  \,,\\
X&:= e^{ 2 b \pi x} \,, \quad\quad Y_i:=-\sqrt{q}\,e^{b \pi u_i }  \,.
\end{align}
The above superpotential is consistent with the well-known fact that the one-loop contribution of each fundamental chiral multiplet $\mathbf{C}$ to $k^{eff}$ is $1/2$, and antifundamental  $\mathbf{AC}$ to $k^{eff}$ is $-1/2$. Moreover,
parity anomaly constrains effective CS levels $k^{eff} \in \mathbb{Z}$. 
The mirror transformation $\mathbf{(1,1, \dots, 1)}$ replaces double sine function $s_b(\dots)$ given by chiral multiplets into contour integrals via \eqref{mirroridentity}.
Hence we get the sphere partition functions for the dual $\T_{A,N}$ theories on the right-hand side of \eqref{U1NtoTAN},
\begin{align}\label{STofT_{1,N}}
Z_{S_b^3}^{  \T_{A,N}}&=
 \int \prod\limits_{i=1}^Nd\, y_i\, e^{\sum\limits_{ i,j=1}^{N} -\pi \,i\, \tilde{k}_{ij} \,y_i y_j+ 2 \pi \, i\, \tilde{\xi}_i y_i    }
\prod\limits_{i=1}^N s_b\big(  \frac{i Q}{2} -y_i  \big) \,,\\
  \tilde{k}_{ij}&=\frac{1}{2}\delta_{ij}-\frac{2}{ 2 k +N}  \,, \nn\\
 \tilde{\xi}_i&= \frac{ i Q}{4}+\frac{u_i}{2}-\frac{2}{2k+N}
\Big( \xi- \sum\limits_{i=1}^{N}
\big(  
\frac{i Q}{4} +\frac{u_i}{4}
\big)   \Big) \nn \,,
\end{align}
where mass parameters $u_i$ can also be absorbed into new FI parameters $\txi_i$. 
When $k=-N/2$, Eq. \eqref{STofT_{1,N}} is ill defined because there is a pole in $\txi_i$, and hence quiver reductions appear in this case. We will show in examples in section \ref{tongpair} that when  $k=-N/2$, this pole can be bypassed and it gives rise to
the mirror pair discovered by Dorey and Tong in \cite{Dorey:1999rb,Tong:2000ky}. In addition, quiver reduction always reduces $(1)_k-[N]$ to a bunch of chiral multiplets after the mirror transformation $(\2, \2 \dots,\2)$ on \eqref{STofT_{1,N}}. However, this involves subtle issues that require taking into account superpotentials for chiral multiples.% so we leave this for future work.

Once we constructed some particular $\T_{A,N}$ theory with the mixed CS level in \eqref{STofT_{1,N}}, acting on it with mirror transformations could lead to many equivalent mirror dual theories. After ruling out theories  with parity anomaly, we can find many equivalent sets of mixed CS levels.% We will provide explicit examples of such results in what follows.

\subsection{Vortex partition functions}\label{vortexquiver}

The correspondence \eqref{equalence} can be independently conjectured (rather than derived) from vortex partition functions, by invoking mathematical identities. In this section we explain this statement taking advantage of the quiver structure found in \cite{Panfil:2018faz, Cheng:2021aa}.

Using the topological vertex formalism for the toric diagram shown in figure \ref{fig:strip},  the vortex partition functions of $U(1)_k+N_C\mathbf{C}+N_{AC} \mathbf{AC}$ theory can be written in the form
%\footnote{The $j$ of $(\b_j,q)_n$ goes from 1 to $N_C-1$, because the parameter $z$ is also for a fundamental chiral $\mathbf{C}$. }
\begin{align}
\label{vortexNCNAC}
Z^{\text{vortex}}_ {U(1)_k+N_C\mathbf{C}+N_{AC} \mathbf{AC}}=\sum\limits_{n=0}^{\inf} \frac{  (-\sqrt{q})^{(f+1)\, n^2}  \big( q^{-\frac{f+1}{2}} z \big)^n}{(q,q)_n} 
\frac{
	\left(\a_1 ,q \right)_n \left(\a_2,q\right)_n  \dots \left(\a_{N_{AC}},q \right)_n 
}
{ 	\left(\b_1 ,q \right)_n \left(\b_2  ,q \right)_n \dots \left(\b_{N_C-1} ,q \right)_n 	
}   \,, 
\end{align}
where $f$ is the framing number that can be put in by band, and the factor $q^{-(f+1)/2}$ can be absorbed into $z$ (see \cite{Kozcaz:2010af, Panfil:2018faz, Cheng:2021aa} for more details). In open topological string theory, open strings are given by M2-branes wrapping a chain of $\mathbb{CP}^1$'s connected to a disk. In terms of refined GV formula \eqref{openGV}, each open string has K\"ahler parameter
\begin{align}
e^{-R \, T_\mathcal{C}}=z^n \prod_{i=1}^{N_{AC}} \prod_{j=1}^{N_C-1}\a_i^{d_i}   \b_j^{d_j}  \,,
\end{align}
where $(n,d_i, d_j)$ are degrees for $(z, \a_i,b_j)$, $z$ is the open K\"ahler parameter for the disk, and 
$\a_i, \b_j$ are closed K\"ahler parameters  from $\mathbf{AC}$ and $\mathbf{C}$, respectively, which correspond to the distances between D5-branes as shown in figure \ref{fig:strip}. The computation reveals that closed K\"ahler parameters correspond to mass parameters of chirals $\a_i, \b_i \sim e^{b \pi u_i}$.

The open topological string partition function \eqref{vortexNCNAC} can be written in the quiver form \cite{Panfil:2018faz, Cheng:2021aa}\footnote{This quiver form comes from quiver representation theory.}
\begin{align}\label{vortextoPc}
Z^{\text{vortex}}_ {U(1)_k+N_C\mathbf{C}+N_{AC} \mathbf{AC}} = Z_{0} \cdot P_{C}({x_1, \dots, x_m})  \,,
\end{align}
where 
\begin{align}
	Z_0=\frac{
		\left(\a_1 ,q \right)_\inf \left(\a_2,q\right)_\inf  \cdots \left(\a_{N_{AC}},q \right)_\inf 
	}
	{ 	\left(\b_1 ,q \right)_\inf \left(\b_2  ,q \right)_\inf \cdots \left(\b_{N_C-1} ,q \right)_\inf	
	} 
	\end{align}
and $P_{C}(\dots)$ is defined as
\begin{align} \label{quivergenfun}
 P_{C}({x_1, \dots, x_m} ):=\sum_{  d_1,...,d_m=0}^{\inf}  
(-\sqrt{q})^{ \sum\limits_{i,j=1}^m C_{ij}d_i d_j} 
\frac{
x_1^{d_1}	x_2^{d_2}\cdots x_m^{d_m}
} 
{ 
(q,q)_{d_1}	(q,q)_{d_2}\cdots (q,q)_{d_m} }
\,,
\end{align}
which is determined by matrices $C_{ij}$.  In \eqref{quivergenfun},  $n$ is denoted by $d_1$ for convenience. To get the form \eqref{quivergenfun} we use the following expansion formula to rewrite each Pochhammer symbol in \eqref{vortexNCNAC}
\begin{align}\label{Cijforaijpm}
(\a_i ,q )_n^{\pm}  &\sim \sum_{d_i=0}^\inf  (-\sqrt{q})^{C_{0,0}[\a_i]n^2+2C_{0,i} [\a_i] n d_i +C_{ii} [\a_i]d_i^2 } \frac{ x_i^{d_i} }{(q,q)_n} \, ,
\end{align}
where $C_{\cdot\cdot}[\a_i]$ denotes the coefficients in front of the degrees $n, \,d_i$. These $C_{\cdot\cdot}[\a_i]$'s encode the presence of chiral multiplets.
Interestingly, there are two equivalent ways to expand Pochhammer symbols, in either $\a_i$ or $\sqrt{q}\a_i^{-1}$, $\b_j$ or $q\, \b_j^{-1}$:
\begin{align}
\label{factors1}
 (\a_i;q)_n &= \frac{ (\a_i ,q)_\inf    }{(\a_i q^n,q)_\inf   } =(\a_i ,q)_\inf   \,
\sum_{d_i=0}^{\inf} (-\sqrt{q})^{
 2  \, n d_i 
} \frac{{\a}_i^{d_i}}{ (q;q)_{d_i} }  \\
&= (\a_i ,q)_\inf   \, (\a_i/\sqrt{q})^{n} 
\sum_{d_i=0}^{\inf} (-\sqrt{q})^{
	n^2 -2  \, n d_i +d_i^2 
} \frac{(\sqrt{q} \a^{-1}_i)^{d_i}}{ (q;q)_{d_i} }  
  \,, \\
  \label{factors2}
\frac{1}{ (\b_j;q)}_n   &= \frac{ (\b_j q^n,q)_\inf}{ (\b_j,q)_\inf  } =\frac{ 1}{ (\b_j,q)_\inf  }
\sum_{d_j=0}^{\inf} (-\sqrt{q})^{ 2n d_j +d_j^2 } \frac{ \left( \frac{{\b}_j}{\sqrt{q}}  \right)^{d_j}}{ (q;q)_{d_i} } \\
&=\frac{ 1}{ (\b_j,q)_\inf  }  \, (  \sqrt{q}/\b_j )^n
\sum_{d_j=0}^{\inf} (-\sqrt{q})^{ -n^2-2 n d_j  } \frac{  \left({  q \, \b_j^{-1} } \right)^{d_j}}{ (q;q)_{d_i} } 
   \,.   
\end{align}
Following this notation, we denote the expansion \eqref{Cijforaijpm} by
% to show only the component $C_{\cdot \cdot}[\a_i]$ and variable $x_i$
\begin{align}
(\a_i,q )^{\pm}_n \rightarrow (    
\left[
\begin{array}{ccc}
C_{0,0}[\a_i] & \cdots&C_{0,i}[\a_i]	  \\
\vdots&\ddots &\vdots\\
C_{i,0}[\a_i]&\cdots&C_{i,i}[\a_i]
\end{array}
\right] , ~x_i
)
\end{align}	
so that each antifundamental chiral $\mathbf{AC}$ leads to\footnote{Where the number marked in blue stands for $C_{0,0}$, which is the open K\"ahler parameter $z$.}
\begin{align}\label{atoCij}
&(\a_i;q)_n \rightarrow
\big(
\left[
\begin{array}{ccc}
\textcolor{blue}{0} &\cdots	  &1 \\
\vdots& \ddots &\vdots \\
1 &\cdots  &0 \\
\end{array}
\right] 
\,,~ \a_i \big) 
\,,  \quad \text{or}\quad
\big(        
\left[ 
\begin{array}{ccc}
\textcolor{blue}{1} &\cdots  &-1 \\
\vdots& \ddots &\vdots\\
-1 &\cdots  &1\\
\end{array}
\right] \,,~ \sqrt{q}\,\a_i^{-1}
\big)\,, &  
\end{align}
and each fundamental chiral $\mathbf{C}$ leads to 
\begin{align}\label{btoCij}
&\frac{1}{ (\b_j;q)}_n \rightarrow
\big(
\left[ 
\begin{array}{ccc}
\textcolor{blue}{0} &\cdots  &1 \\
\vdots& \ddots &\vdots\\
1 &\cdots  &1\\
\end{array}
\right] \,,~
\frac{\b_j }{ \sqrt{q}}
\big)
\,,   \quad \text{or } \quad
\big(  
\left[ 
\begin{array}{ccc}
\textcolor{blue}{-1} &\cdots  &-1 \\
\vdots& \ddots &\vdots\\
-1 &\cdots  &0\\
\end{array}
\right] \,,~ \sqrt{q}\Big(\frac{\b_j}{\sqrt{q}}\Big)^{-1}
\big)  \,, 
&        
\end{align}
where all the elements denoted by ``$\dots$" in the above matrices are  $0$. In total, 
the matrix $C_{ij}$ has the structure
\begin{align}
C_{ij}=C_{\cdot  \cdot}[z]+\sum_i C_{\cdot \cdot}[\a_i]+\sum_jC_{\cdot \cdot}[\b_j]  \,.
\end{align}
Here we show one particular CS level matrices $k_{ij}^{eff}$ for  \eqref{vortexNCNAC}: fixing 
%by using \eqref{atoCij}  and \eqref{btoCij}. 
the variables $x_i$ in $P_C(\dots)$ as follows:
\begin{align}
 P_C(x_0, x_1\,,\dots,x_{m} )=P_C\left(  q^{-\frac{f+1}{2}}z, \a_1 \,,  \dots\,, \a_{N_{AC}} \,,~ \frac{\b_1  }{ \sqrt{q}}\,,\dots\,,  \frac{\b_{N_C-1} }{ \sqrt{q} } \right)  \,,
\end{align}
we find that the  $C_{ij}$ matrix takes form
\begin{align}
C_{ij} ( \eqref{vortextoPc})= \left[\begin{array}{c|ccc|ccc}
\color{blue}{f+1}& 1 & \dots &  1 & 1 & \dots & 1  \\ \hline
1 & 0 & \dots & 0 & 0 & \dots & 0 \\
\vdots & & \ddots & & & \ddots &  \\
1 & 0 & \dots & 0 & 0 & \dots & 0 \\ \hline
1 & 0 & \dots & 0 & 1 & \dots & 0 \\
\vdots & & \ddots & & & \ddots &  \\
1 & 0 & \dots & 0 & 0 & \dots & 1 \\
\end{array}\right].
\end{align}
The rank of $C_{ij}$ is $(N_{AC}+N_{C}) \times  (N_{AC}+N_{C})$.
By comparing superpotentials in explicit examples, we find that the framing number is related to the bare CS level $k$,
\begin{align}
f+1=k+\frac{N_C-N_{AC}}{2}  \,.
\end{align}
Note that there are several ways to write $Z^{\text{vortex}}_ {U(1)_k+N_C\mathbf{C}+N_{AC} \mathbf{AC}}$ in the form of $P_C(x_i)$, since there are two equivalent expansion parameters $x_i$
in \eqref{atoCij} and \eqref{btoCij}.
 If
flipping any  $x_i \rightarrow \sqrt{q} \,x_i^{-1}$, then one gets another matrix $C_{ij}'$. All $x_i$ can be flipped, and therefore one gets a chain of $\{C_{ij}\}$. There are in total $2^{ N_{AC}+N_{C}-1}$ equivalent matrices.

Invoking the mirror symmetry,  we can provide a physical interpretation of \eqref{vortextoPc} and matrices $C_{ij}$. Recall that   \eqref{vortextoPc} implies that
the vortex partition functions of $U(1)_k+N_C\mathbf{C}+N_{AC} \mathbf{AC}$ theories can be rewritten in the quiver form $P_{C_{ij}}(x_i)$.  It can be noticed that on the Higgs branch, $Z_0$ is actually related to the one-loop part $ Z^{\text{1-loop}} =Z_0^{-1}$,
which is given by the inverse of Pochhammer symbols in \eqref{factors1}-\eqref{factors2}
\begin{align}
Z^{\text{1-loop}}_ {U(1)_k+N_C\mathbf{C}+N_{AC}}= \frac{  \prod_{ j=1}^{N_{C}}
	\left(\b_j  ,q \right)_\inf }	
{
	\prod_{ i=1}^{N_{AC}}
	\left(\a_i  ,q \right)_\inf
}   \,,
\end{align}
and then    \eqref{vortextoPc} reads
\begin{align}\label{U(1)NftoP}
 Z^{\text{1-loop}}_ {U(1)_k+N_C\mathbf{C}+N_{AC} \mathbf{AC}} \cdot Z^{\text{vortex}}_ {U(1)_k+N_C\mathbf{C}+N_{AC} \mathbf{AC}}(z,\a_i,\b_j ) = P_{C_{ij}}( x_i )  \,.
\end{align}
%see \cite{Pasquetti:2011fj} for some examples on one-loop part.
Moreover, vortex partition functions \eqref{vortexeff} of $\T_{A,N} $ theories also  take a quiver form 
\begin{align}\label{quiverisCS}
Z^{\text{vortex}}_ {\T_{A, N_C+N_{AC}}} (k_{ij}^{eff},x_i  ) = P_{C_{ij}}(x_i) 
\,,
\end{align}
hence we conjecture $C_{ij} =k^{eff}_{ij} $ and
% \eqref{U(1)NftoP} and \eqref{quiverisCS} imply that 
 $U(1)_k+ N_C \,\mathbf{C} + N_{AC} \, \mathbf{AC}$ can be regarded as certain  $\T_{A,N}$ theories. 
Then the vortex partition functions of $U(1)_k+N_C\mathbf{C}+N_{AC} \mathbf{AC}$ theories are conjectured to be equal to vortex partition functions of the corresponding $\T_{A,N}$ theories
\begin{equation}
\boxed{
\label{vortextovortex}
Z^{\text{1-loop}}_{U(1)_k+N_C\mathbf{C}+N_{AC} \mathbf{AC}} (\a_i,\b_j ) \cdot Z^{\text{vortex}}_ {U(1)_k+N_C\mathbf{C}+N_{AC} \mathbf{AC}}(z,\a_i,\b_j ) = Z_ {\T_{A, N_C+N_{AC}}} (x_i  )
 . }
\end{equation}
This is checked to be correct in various examples in the following sections.
We stress that the one-loop part of the $\T_{A,N}$ theory on the Higgs branch is trivial, and hence $Z_ {\T_{A, N_C+N_{AC}}} (x_i  )\backsimeq Z^{\text{vortex}}_ {\T_{A, N_C+N_{AC}}} (x_i  )$. Note that the correspondence between $U(1)_k+N_C\mathbf{C}+N_{AC} \mathbf{AC}$  and $\T_{A,N}$ theories is a conjecture from the perspective of vortex partition functions; however, this correspondence can be derived from the sphere partition functions using mirror transformations.  Furthermore, the vortex partition functions in  \eqref{vortextovortex} can be refined, and then they
satisfy refined open GV formula \eqref{openGV} that encodes positive integer BPS numbers; for more details and explicit computations see \cite{Cheng:2021aa}.

There is one problem left: what are the relations between these equivalent $C_{ij}$'s? The answer is that each $C_{ij}$ is the $k_{ij}^{eff}$ of a particular mirror dual theory, and mirror symmetry relates them. More explicitly,
mirror transformations relate dual theories 
\begin{align}
\T[ \mathbf{( n_1,\dots, n_{N_C+N_{AC}})} ]  \xlongrightarrow{{ \mathbf{( i_1, \dots, i_{N_C+N_{AC}} )  }}}
\T[ \mathbf{( n_1+i_1,\dots, n_{N_C+N_{AC}} +i_{N_C+N_{AC}}   )} ]   \,,
\end{align}
and give rise to mirror maps between effective CS levels
\begin{align}
k_{ij}^{eff, \, \mathbf{( n_1,\dots, n_{N_C+N_{AC}})}  }  \xrightarrow{ \text{flipping some $x_i \rightarrow x_i^{-1}$}}  k_{ij}^{eff,\, \mathbf{( n_1+i_1,\dots, n_{N_C+N_{AC}} +i_{N_C+N_{AC}}   )}    }  \,.
\end{align}
We will show in examples in the following sections that these equivalent integer CS matrices $k_{ij}^{eff}$ can be obtained by performing mirror transformations on sphere partition functions. In terms of vortex partition function of the corresponding  $\T_{A,N_C+N_{AC}}$ theories, mirror symmetry acts as flipping closed K\"ahler parameters $\a_i  \rightarrow \a_i^{-1}$ or $\b_j \rightarrow \b_j^{-1}$(or in other words, changing the sign of real mass parameters $u_i \rightarrow -u_i$, since the closed K\"ahler parameters equal to mass parameters and FI parameters by $
\a_i\,,\b_i \sim e^{\pi b \, u_i}\,, z\sim e^{2 b \pi {\xi}}$).
However, the exchange symmetry $q \rightarrow 1/q$ in open topological strings does not lead to new  CS level matrices $k_{ij}^{eff}$, as it only shifts bare CS level $k \rightarrow k\pm 1$.

\subsection{$U(1)_k+1 \, \mathbf{C}$}

$\T_{k,1}:\,(1)_k-[1]$ theory is an interesting basic example.  Its sphere partition function is given by \eqref{shperepartNC}. We shift $x$ and absorb the mass parameter in $\tilde{\xi}$ and obtain
\begin{align}
Z^{\T_{k,1}}= \int d x \, e^{  2 \pi \, i\, \tilde\xi x- i\, \pi k x^2 } s_b\big(  \frac{i Q}{2}-x \big)  \,.
\end{align}
The mirror transformation group $\mathcal{H}(\T_{A,1})$ in this case takes the form
\begin{align}
\mathcal{H}(\T_{A,1})= \{ \mathbf{(0), (1),(2)} \} \,,
\end{align}
which leads to mirror dual theories \begin{align}\{ ~\T[  \mathbf{(0)}] \,,~\T[  \mathbf{(1)}] \,,  ~\T[  \mathbf{(2)}]    ~	\}\,. \end{align}
Mirror transformation ${\mathbf{(1)}}$ relates them as follows:
\begin{align}
\T[\mathbf{(0)}] \xrightarrow{\mathbf{(1)}} \T[\mathbf{(1)}]  \xrightarrow{\mathbf{(1)}} \T[\mathbf{(2)} ]  \,,
\end{align}
namely,
\begin{align}
&(1)-[1] ~ \xrightarrow{{\mathbf{(1)}}} ~
 (1)-((1)-[1]  ) ~\xrightarrow{{\mathbf{(1)}}}~ (1)-((1)-((1)-[1])  )  \,,
\end{align}
which are the following quivers after integrating out old gauge nodes:
\begin{align}
(&1)_k-[1]  ~\xrightarrow{{\mathbf{(1)}}}~
(1)'_{k'}-[1]  ~\xrightarrow{{\mathbf{(1)}}}~ (1)''_{k''}-[1] \,.
\end{align}
Their sphere partition functions are as follows:
\begin{align}
\begin{split}
Z^{\T[ \mathbf{0}]	}_{S_b^3}&= \int dx\, e^{ 2 \pi \, i\, \txi x  - k \pi \,i\, x^2} s_b \big(   \frac{i Q}{2} -x \big)  \,,\\
Z^{\T[ \mathbf{1}]}_{S_b^3}&= \int dx\, e^{
	\frac{   \pi ( Q-2k Q -8 i \txi )x+i ( 3-2k)\pi x^2     }
	{  2+4 k  }
} s_b \big(   \frac{i Q}{2} -x \big)  \,,\\
Z^{ \T[ \mathbf{2}]  }_{S_b^3}&= \int dx\, e^{
	\frac{   \pi  ( Q+2k Q +8 i \txi )x +i ( 3+2k) \pi x^2     }
	{  -2+4 k  }
} s_b \big(   \frac{i Q}{2} -x \big)  \,.
\end{split}
\end{align}
One can see mirror transformations change CS levels and FI parameters significantly. By taking semi-classical limit and using formula \eqref{WeffQ}, we read off these CS levels and FI parameters
\begin{align}\label{(1)[1]class}
\begin{split}
& 
\T[ \mathbf{0}]: ~	
\big( ~k_{ij}^{eff,\mathbf{( 0)} }=\frac{1}{2}+k\,, ~
~\, \xi^{eff, \mathbf{(0)}} =2 b \pi \txi + i \pi(1-b Q)k +\frac{i \pi}{2}~
\big) \,,\\
&
\T[ \mathbf{1}]: ~	
\big(~ k_{ij}^{eff, \mathbf{( 1)} }=\frac{ 2k-1} {  2k+1 }  \,,~
\xi^{eff, \mathbf{( 1)}} =- \frac{4 b \pi   \txi }{ 1+2 k} 
+\frac{     i \pi(2 k  -1+b Q)       }{1+2k}
~\big) \,,  \\
&
\T[ \mathbf{2}]: ~	
\big( ~ k_{ij}^{eff,\mathbf{( 2)}}=\frac{ 2} {  1-2k }  \,,~
\xi^{eff,\mathbf{(2)}} = \frac{ i(-2 \pi +b \pi Q    -4 i b \pi \txi)}{2k-1}
~\big) 
\,.
%\}\,.
\end{split}
\end{align}
As we discussed before, mirror transformations  permute mirror dual theories.  
The permutation 
\begin{align}
 \T[\mathbf{(0)}]\rightarrow  \T[\mathbf{( 1)} ] , ~\T[\mathbf{( 1)} ] \rightarrow \T[\mathbf{( 2)} ] ,~ \T[\mathbf{( 2)}] \rightarrow \T[ \mathbf{( 0)} ] 
\end{align}
 is given by mirror transformation $\mathbf{( 1)}$, and the corresponding mirror map is
\begin{align}\label{mirrorfor(1)[1]}
(k,\txi) \rightarrow (k', \txi') :~
k'= \frac{ 3+2 k}{2- 4 k} \,,~ \txi'=\frac{ i(Q+2 k Q +8 i \txi  )  }{ 4 -8 k  }  \,.
\end{align}
The permutation given by mirror transformation $\mathbf{(2)}$ is
\begin{align}
\T [\mathbf{(0)}] \rightarrow \T[\mathbf{( 2)}],~ 
	\T[\mathbf{(2)} ] \rightarrow \T[\mathbf{( 1)} ] ,~
 \T[\mathbf{( 1)} ] \rightarrow \T[\mathbf{(0)}] \,
\end{align}
whose corresponding mirror map is the reverse of \eqref{mirrorfor(1)[1]}
\begin{align}\label{mirrorfor(1)[1]2}
(k,\txi) \rightarrow (k'', \txi'') :~
k''= \frac{ -3+2 k}{2+4 k} \,,~ \txi''=\frac{ i(-1+2 k) Q -8  \txi    }{ 4 +8 k  }  \,.
\end{align}
In this paper, we only consider mirror maps for $\T_{A,1}$ theories. In principle,
one can find mirror maps for generic $\T_{A,N}$ theories too.

Parity
anomaly constrains $k_{ij}^{eff,( i)}$ to be integers, so we throw away theories with fractional effective CS levels, and find all possible values for bare CS level $k$ 
%that  can lead to mirror dual pairs
\begin{align}
k=\pm 3/2,0,\pm 1/2\,. 
\end{align}
The associated effective CS levels and FI parameters can be obtained by inserting these values in \eqref{(1)[1]class}.
%When $k=\pm 3/2,0$, we take $k$ into \eqref{(1)[1]class} and obtain several theories denoted by $\T_\bullet$:

More explicitly, when $k=\pm 3/2$, we get theories denoted by $\T_{1,2,3,4,5,6}$,
\begin{align}
&k=-\frac{3}{2}, \quad  \T_1 \xrightarrow{{\mathbf{( 1)}}} \T_2  \,, \\
& k=~~0,~ \quad \T_3 \xrightarrow{{\mathbf{( 1)}}} \T_4  \,,  \\
&k=~~\frac{3}{2}, \quad  \T_5 \xrightarrow{{\mathbf{( 2)}}} \T_6		  \,, 
\end{align}
where
\begin{align}
\begin{split}
&\T_1: ~\{ k_{ij}^{eff, \mathbf{( 0)}}=-1  \,, ~ \xi^{eff, \mathbf{(0)}}=-i \pi + \frac{3}{2} i \,b \pi Q + 2\, b \pi \txi  \}   \\
&  \T_2 : ~\{  k_{ij}^{eff, \mathbf{( 1)}}=2 \,,  ~  ~~\, \xi^{eff, \mathbf{( 1)}}=2\, i \pi - \frac{1}{2} i \,b \pi Q + 2\, b \pi \txi   \} \,,\\
&\T_3 : ~ \{ k_{ij}^{eff, \mathbf{( 1)}}=-1 \,,~   \xi^{eff, \mathbf{( 1)}}=-i \pi + i \,b \pi Q -4 \, b \pi \txi    \} \,, \\
& \T_4 : ~\{  k_{ij}^{eff, \mathbf{( 2)}}=2 \,,~~~\, \xi^{eff, \mathbf{( 2)}}=2\, i \pi -i \,b \pi Q -4\, b \pi \txi      \} \,,\\
&  \T_5 : ~ \{ k_{ij}^{eff, \mathbf{(0)}}=2 \,,\,\quad \xi^{eff, \mathbf{(0)}}=2 \,i \pi - \frac{3}{2} i \,b \pi Q + 2 b \pi \txi   
\} \,,   \\
&  \T_6 : ~  \{  k_{ij}^{eff, \mathbf{(2)}}=-1 \,, ~\,   \xi^{eff, \mathbf{( 2)}}=-i \pi + \frac{1}{2} i \,b \pi Q + 2 b \pi \txi  \} \,. 
\end{split}
\end{align}
Some of them are equivalent
\begin{align}
&\T_1 = \T_3=\T_6:~(1)_{-3/2}-[1] \,, \\
& \T_2  = \T_4 = \T_5:~ (1)_{3/2}-[1] \,.
\end{align}
Therefore, we end up with a mirror dual pair
\begin{align}
\{ ~(1)_{ 3/2}-[1]  \,, ~~(1)_{ -3/2}-[1]  ~\} \,.
\end{align}

When $k=\pm 1/2$,  and inserting this value into \eqref{(1)[1]class}, we find the theories
\begin{align}
&k=~~\frac{1}{2}, \quad\quad  \T_7 \xrightarrow{{\mathbf{( 1)}}} \T_8 \xrightarrow{{\mathbf{( 1)}}} \T_9  \,, \\
&k=-\frac{1}{2}, \quad\quad \T_{10} \xrightarrow{{\mathbf{( 1)}}} \T_{11} \xrightarrow{{\mathbf{( 1)}}} \T_{12} 	  \,, 
\end{align}
where
\begin{align}
\begin{split}
& \T_7:  \{ k_{ij}^{eff,  \mathbf{(0)}}=1, ~ \xi^{eff, \mathbf{(0)}}=i \pi -\frac{1}{2} i\, b\pi  Q + 2 b \pi \txi  \,  \}   \\
&  \T_8:  \{  k_{ij}^{eff,  \mathbf{(1)}}=0 \,, \xi^{eff,(\1)}=\frac{1}{2} i\, b\pi  Q - 2 b \pi \txi  \,      \}  \\
& \T_9:  \{  k_{ij}^{eff, (\2)}=\inf  \}     \label{mirrorind1}   \,,  \\
& \T_{10}: \{ k_{ij}^{eff, (\0)}=0  \,, \xi^{eff,(\0)}=\frac{1}{2} i\, b\pi  Q +2 b \pi \txi  \,     \}  \\
& \T_{11}: \{  k_{ij}^{eff, (\1)}=\inf  \}  \\
&
\T_{12}:
\{  k_{ij}^{eff, (\2)}=1 \,,
\xi^{eff,(\2)}=i \pi -\frac{1}{2} i\, b\pi  Q - 2 b \pi \txi  \, 
\}     \,.  
\end{split}
\end{align}
Here $\inf$ implies that there is a quiver reduction.
%The above theories are related by $\mathbf{(1)}$
Moreover, some of these theories are equivalent
\begin{align}
 \T_7 =  \T_{12} \,, ~~ \T_8= \T_{10}  \,,~~ T_9=\T_{11} \,.
\end{align}
 More explicitly, when $k=1/2$, sphere partition functions for $\T_{7,8,9}$ take the form (where we define $\txi:=  -p + \frac{i Q}{4}$)
\begin{align}
& \T_7: ~ Z^{(1)_{1/2}-[1]}_{S_b^3}~\,= \int dx\, e^{ 2 \pi \, i\,\big( \frac{i Q}{4}-p \big)  x  - \frac{1}{2} \pi \,i\, x^2} s_b \big(   \frac{i Q}{2} -x \big)  \,,\\
& \T_8 :~  Z^{(1)_{-1/2}-[1]}_{S_b^3}= \int dx\, e^{ -2 \pi \, i\,\big( \frac{i Q}{4}-p \big)  x  + \frac{1}{2} \pi \,i\, x^2} s_b \big(   \frac{i Q}{2} -x \big)  \,,\\
& \T_9:~   Z^{[1]_{-1/2}-[1]}_{S_b^3}\,= e^{   
	\frac{ i \pi}{ 2} \big( \frac{i Q}{2} -p   \big)^2} s_b\big( \frac{i Q}{2} -p\big) \,.
\end{align}
The mirror transformations relate these three theories, and hence
\begin{align}\label{pair1}
Z^{(1)_{1/2}-[1]}_{S_b^3}=Z^{(1)_{-1/2}-[1]}_{S_b^3}=Z^{[1]_{-1/2}-[1]}_{S_b^3} 
\end{align}
where $Z^{(1)_{1/2}-[1]}_{S_b^3}=Z^{[1]_{-1/2}-[1]}_{S_b^3}  $ is the identity in \eqref{oldms}.
Similarly, when $k=-1/2$, partition functions for $\T_{10,11,12}$ are of the following form (where we define $\txi := p - \frac{i Q}{4}$)
\begin{align}
& \T_{10}: ~ Z^{(1)_{-1/2}-[1]}_{S_b^3}= \int dx\, e^{ -2 \pi \, i\,\big( \frac{iQ}{4}-p \big)  x  + \frac{1}{2} \pi \,i\, x^2} s_b \big(   \frac{i Q}{2} -x \big)  \,,\\
& \T_{11}:~ Z^{[1]_{1/2}-[1]}_{S_b^3}~~= e^{   
	-\frac{ i \pi}{ 2} \big( \frac{i Q}{2} -p   \big)^2} s_b\big( \frac{i Q}{2} -p\big)  \,, \\
& \T_{12}:~ Z^{(1)_{1/2}-[1]}_{S_b^3}~\,= \int dx\, e^{ 2 \pi \, i\,\big( \frac{i Q}{4}-p \big)  x  - \frac{1}{2} \pi \,i\, x^2} s_b \big(   \frac{i Q}{2} -x \big)  \,.
\end{align}
The mirror transformation relates them as follows:
\begin{align}\label{pair2}
Z^{(1)_{-1/2}-[1]}_{S_b^3}=Z^{[1]_{1/2}-[1]}_{S_b^3}=Z^{[1]_{1/2}-[1]}_{S_b^3}  \,.
\end{align}
Combining \eqref{pair1} and \eqref{pair2}, we get another mirror pair
\begin{align}
\{ ~ (1)_{ 1/2}-[1],~~ (1)_{ -1/2}-[1],~~ [1]_{ 1/2}-[1] ,~~[1]_{- 1/2}-[1] ~  \} \,.
\end{align}
\begin{figure}
	\centering
	\includegraphics[width=1.2in]{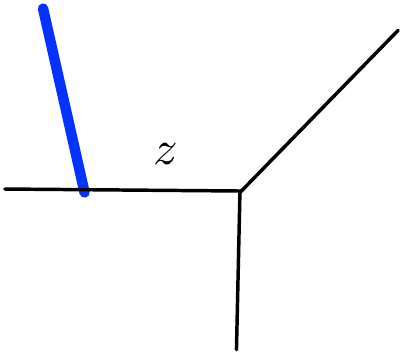} 
	\caption{Calabi-Yau threefold $\mathbb{C}^3$ with a Lagrangian brane marked in blue.}
	\label{fig:C^3braneweb}
\end{figure}

The toric diagram for the theory $(1)_{k}-[1]$ is shown in figure \ref{fig:C^3braneweb}. 
By \eqref{vortexNCNAC}, the open K\"ahler parameter for the open topological  brane on Calabi-Yau threefold $\mathbb{C}^3$ is $q^{(f+1)/2} z$ where $f $ is the framing number.
To match it with the FI parameter in vortex partition functions in \eqref{vortexeff}, we identify
\begin{align}
e^{\xi^{eff, \mathbf{(0)}}} =i\,(-1)^{k}   q^{-\frac{k}{2}}  e^{2 b \pi \txi  } =(-1)^{f+1} q^{ -\frac{f+1}{2}} z
\end{align}
which implies that the framing number $f$ maps to the CS level $k$, and the open  K\"ahler parameter maps to the FI parameter
\begin{align}\label{relationC^3}
f=k-1/2\,,\quad
z=  q^{1/4} \, e^{2 b \pi \txi  } \,.
\end{align}

\subsection{$U(1)_k+ 2\, \C$}
We turn this theory into a particular 	$\T_{A,N}$ theory
\begin{align}
Z_{S_b^3}^{(1)_{k}+ 2\mathbf{C} }
\xlongrightarrow{\mathbf{(1,1)}}
Z_{S_b^3}^{\T_{A,2}  }
\end{align}
where $Z_{S_b^3}^{\T_{A,2}  }$ is given by 
 \eqref{STofT_{1,N}} when $N=2$. 
We perform mirror transformations $\mathbf{(n_1,n_2)} \in \H(\T_{A,2})$ and take the semiclassical limit to read off effective superpotentials. 
For simplicity, we denote the mirror dual theories and superpotentials by $\T[\mathbf{(n_1,n_2)}]:   (k_i^{eff, \mathbf{(n_1,n_2)}},\, \xi_i^{eff,\mathbf{(n_1,n_2)}} )$ and find the following results 
\begin{align}\label{data2C}
\begin{split}
&\T[\mathbf{(0,0)}]:~
\left(
\begin{pmatrix}
\frac{k}{1+k} & -\frac{1}{1+k}  \\
- \frac{1}{1+k} & \frac{k}{1+k}   \\
\end{pmatrix} \,, ~
\begin{pmatrix}
\frac{\pi (2 i (k-1+ bQ-2 i b \xi  )+( b+2 b k)u_1 - b u_2  )}{2(1+k)}  \\
\frac{\pi (2 i (k-1+ bQ-2 i b \xi  ) -b u_1 +(b+2 b k) u_2     }{2(1+k)}   \\
\end{pmatrix}   \right)
\,,  \\
&\T[\mathbf{(0,1)}]:~
\left(
\begin{pmatrix}
\frac{k-1}{k}& \frac{1}{k}  \\
\frac{1}{k}  & -\frac{1}{k}    \\
\end{pmatrix} \,, ~
\begin{pmatrix}
\frac{\pi (2 i k+ 4 b \xi+ b(2k-1)u_1+ b u_2  )}{2 k}   \\
-\frac{b \pi (   4 \xi -u_1 +( 1+2 k) u_2  )}{2 k}      \\
\end{pmatrix} \right) \,,  \\
&\T[\mathbf{(0,2)}]:~
\left(   
\begin{pmatrix}
1 & 1  \\
1 & 1+k  \\
\end{pmatrix} \,, ~
\begin{pmatrix}
\pi (2 i-i b Q+bu_1-b u_2)  \\
\frac{1}{2} \pi  \left(-2 i (b k Q-2 i b \xi +b Q-k-2)-u_2 (2 b
k+b)+b u_1\right) \\
\end{pmatrix}   \right)	   \,, \\
& \T[\mathbf{(1,0)}]:~
\left(
\begin{pmatrix}
-\frac{1}{k} & \frac{1}{k} \\
\frac{1}{k} & \frac{k-1}{k} \\
\end{pmatrix} \,, ~
\begin{pmatrix}
-\frac{\pi  b \left((2 k+1) u_1+4 \xi -u_2\right)}{2 k}  \\
\frac{\pi  \left(b (2 k-1) u_2+4 b \xi +b u_1+2 i k\right)}{2 k}   \\
\end{pmatrix}       \right)
\,, \\
&  \T[\mathbf{(1,1)}]:~
\left( 
\begin{pmatrix}
\frac{1}{1-k} & \frac{1}{1-k} \\
\frac{1}{1-k} & \frac{1}{1-k} \\
\end{pmatrix} \,, ~
\begin{pmatrix}
\frac{\pi  \left(u_1 (b-2 b k)-4 b \xi +2 i b Q-b u_2-4 i\right)}{2 (k-1)} \\
-\frac{\pi  \left(b (2 k-1) u_2+4 b \xi -2 i b Q+b u_1+4 i\right)}{2 (k-1)} \\
\end{pmatrix}      \right)
 \,, \\
& \T[\mathbf{(1,2)}]:~
\left(  
\begin{pmatrix}
0 &- 1  \\
-1 & k  \\
\end{pmatrix} \,, ~
\begin{pmatrix}
\pi  \left(i (b Q-1)-b u_1+b u_2\right) \\
\frac{1}{2} \pi  \left(-2 i (k (b Q-1)-b (Q+2 i \xi )+1)+u_2 (b-2 b k)-b u_1\right)   \\
\end{pmatrix}     \right)
 \,,  \\
&  \T[\mathbf{(2,0)}]:~
\left(  
\begin{pmatrix}
k+1& 1  \\
1 & 1  \\
\end{pmatrix} \,, ~
\begin{pmatrix}
\frac{1}{2} \pi  \left(-2 i (b k Q-2 i b \xi +b Q-k-2)-u_1 (2 b k+b)+b u_2\right)   \\
\pi  \left(-i b Q-b u_1+b u_2+2 i\right) \\
\end{pmatrix}     \right)
\,,  \\
& \T[\mathbf{(2,1)}]:~
\left(  
\begin{pmatrix}
k &- 1  \\
-1 & 0 \\
\end{pmatrix} \,, ~
\begin{pmatrix}
\frac{1}{2} \pi  \left(-2 i (k (b Q-1)-b (Q+2 i \xi )+1)+u_1 (b-2 b k)-b u_2\right)  \\
\pi  \left(i b Q+b u_1-b u_2-i\right) \\
\end{pmatrix}      \right)  \,.
\end{split}
\end{align}
Because of the exchange equivalence $\mathbf{n_i \leftrightarrow n_j}$, there are only two independent theories. We identify these theories with $\{\, \T[\mathbf{(2,0)} ]\,,~ \T[\mathbf{(2,1)} ] \,\}$, which are related by the transformation $\mathbf{(0,1)}$
\begin{equation}\label{mirror2c}
\T[\mathbf{(2,0)} ] \xlongrightarrow{\mathbf{(0,1)} } \T[\mathbf{(2,1)}  \,.
\end{equation}
\begin{figure}
	\centering
	\includegraphics[width=1.2in]{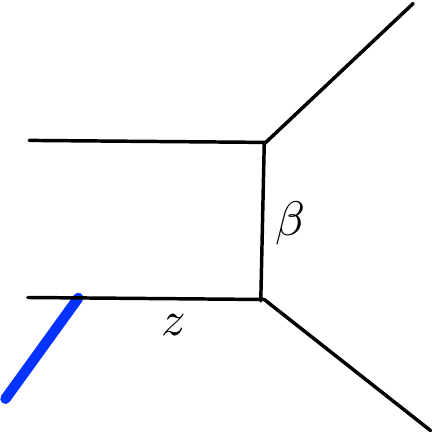} 
	\caption{The toric Calabi-Yau threefold with a Lagrangian brane for  theory $U(1)_k+ 2 \mathbf{C}$. }
	\label{fig:cp1}
\end{figure}
The toric diagram  for the theory $(1)_k-[2]$  is shown in figure \ref{fig:cp1}.
It follows from \eqref{vortexNCNAC} that the vortex partition function takes the form
\begin{align}
\label{vortexU(1)2C1}
Z^{\text{vortex}} _{ U(1)_k+ 2 \,\mathbf{C}   } &=\sum\limits_{n=0}^{\inf} \frac{  (-\sqrt{q})^{(f+1)\, n^2}  ( q^{-\frac{ f +1}{2} } z)^n}{(q,q)_n} 
\frac{1}
{ 	\left(\b,q \right)_n 
}    
\end{align}
which, combined with the one-loop part, takes the form of the vortex partition function of the $\T_{A,2}$ theory. However, there are several equivalent forms of \eqref{vortexU(1)2C1}, as we discussed in section \ref{vortexquiver}, and each form corresponds to the vortex partition function of a particular $\T_{A,2}$ theory
\begin{align}
Z _{ \T_{A,2}}^{\text{vortex}}
&=  Z^{\text{1-loop}} _{ U(1)_k+ 2 \,\mathbf{C}   }     \cdot Z^{\text{vortex}} _{ U(1)_k+ 2 \,\mathbf{C}   }   \label{TaN2C}   \\
&= 
\sum_{ d_1,d_2=0}^{\inf}  
(-\sqrt{q})^{ \sum\limits_{i,j=1}^2 k^{eff,  \mathbf{(2,0)}}_{ij}d_i d_j} 
\frac{
	z^{d_1} \, ( \b/\sqrt{q})^{d_2}
} 
{ 
	(q,q)_{d_1} (q,q)_{d_2} }  \label{2cTAN1}
\\
&= \sum_{ d_1,d_2=0}^{\inf}  
(-\sqrt{q})^{ \sum\limits_{i,j=1}^2 k^{eff,  \mathbf{(2,1)}}_{ij}d_i d_j} 
\frac{
	z^{d_1} \, (q \b^{-1} )^{d_2}
} 
{ 
	(q,q)_{d_1} (q,q)_{d_2} }  \,,  \label{2cTAN2}
\end{align}
where we have absorbed the additional framing number and factors caused by flipping $\b$ into $z$.
From \eqref{mirror2c}, it can be noticed that \eqref{2cTAN1} is the vortex partition function for theory $\T[  \mathbf{(2,0)}]$,  \eqref{2cTAN2} is the vortex partition functions for theory $\T[  \mathbf{(2,1)}]$, and
flipping mass parameter $\b/\sqrt{q} \rightarrow q \b^{-1}$ relates effective CS levels
\begin{equation}\label{keffmirror2c}
\begin{tikzcd}[column sep=3.5pc]
k_{ij}^{eff,  \mathbf{(2,0)} } \arrow{r}{\text{flip}~ \b} & k_{ij}^{eff,  \mathbf{(2,1)} }  
\end{tikzcd}  \,.
\end{equation}
This flipping is interpreted as mirror transformation $\mathbf{(0,1)}$, as $\mathbf{(2,0)+(0,1)=(2,1)}$.  
%In order to match \eqref{vortexeff} with \eqref{TaN2C}, the factor $\sqrt{q}$ and $q$ need to be absorbed into $\b$. This kind of shift also exists for other examples in this note.
	
The relations between K\"ahler parameters $z\,, \a_i\,, \b_j$ and gauge theory parameters $u_i, \xi$ can be obtained by comparing with \eqref{vortexeff} where the variables $x_i$ are defined to be $x_i :=(-1)^{k^{eff}_{ii}} e^{\xi_i^{eff}} $.
For $\T[\mathbf{(2,0)}]$, the relations between K\"ahler parameters and gauge theory parameters are given by
\begin{align}
\big(  q^{-\frac{ f +1}{2} } z,\, \beta/\sqrt{q} \big)=
\left(  (-1)^{2k+1} q^{- \frac{ k+1}{2}} e^{ - b \pi u_1 (\frac{1}{2}+k)}  e^{b \pi u_2/2} e^{-2 b \pi \xi}    \,,  -e^{b \pi (u_2-u_1) }/\sqrt{q} \right) \,,
\end{align}
while for $\T[\mathbf{(2,1)}]$ the relations are 
\begin{align}
\big(  q^{-\frac{ f +1}{2} } z \,, q \beta^{-1} \big)=\left(  (-1)^{k+1} q^{- \frac{ k-1}{2}} e^{ b \pi u_1 (\frac{1}{2}-k)}  e^{-b \pi u_2/2} e^{-2 b \pi \xi}    \,,  -\sqrt{q}  \, e^{b \pi (u_1-u_2) } \right)  \,.
\end{align}
 If $u_1=0$, the relations between $z, \b$ and $u_i, \xi$ simplify to $z \sim  e^{2 b \pi \xi}$ and $\b\sim e^{b\pi u_2}$.

\subsection{$U(1)_k+ 1 \,\mathbf{C}+1\, \mathbf{AC}$}
The sphere partition function for this theory is
\begin{align}
&Z_{S_b^3}^{(1)_{k}+ 1\mathbf{C}+ 1 \mathbf{AS} } = \int d x\, e^{2 \pi \xi x -i \pi k x^2 } s_b\big( \frac{ i Q}{2} +x +\frac{u_1}{2}   \big)
s_b\big( \frac{ i Q}{2} -x +\frac{u_2}{2}   \big)
\end{align}
which after the mirror transformation $\mathbf{(1,1)}$ becomes that of the theory $\T_{A,2}$
\begin{align}
Z_{S_b^3}^{(1)_{k}+ 1\mathbf{C}+ 1 \mathbf{AS} }
\xlongrightarrow{\mathbf{(1,1)}}
Z_{S_b^3}^{\T_{A,2}  }\,,
\end{align}
where
\begin{align}\label{(1)_k+1C+1ASpart}
Z_{S_b^3}^{\T_{A,2}  }= \int & d y_1 dy_2 \,
e^{-i \pi  \frac{k-1 }{k+1} \left(y_1^2 + y_2^2\right)
	-\frac{i \pi  \left(-i k Q-(2 k+1) u_1-4 \xi -i Q-u_2\right)}{2 (k+1)} y_1
	-\frac{i \pi  \left(-i k Q-(2 k+1) u_2+4 \xi -i Q-u_1\right)}{2 (k+1)} y_2
}   \nn\\
&\times s_b\big( \frac{i Q}{2}  -y_1 \big)  s_b\big( \frac{i Q}{2}  -y_2   \big)   \,.
\end{align}
After acting with mirror transformations from the group $\H( \T_{A,2}  )$, we obtain mirror dual theories labeled as follows
\begin{align}
&\T[\mathbf{(0,0)}]:  \left(
\begin{pmatrix}
\frac{k}{1+k} & \frac{1}{1+k}  \\
\frac{1}{1+k} & \frac{k}{1+k}   \\
\end{pmatrix} \,, ~
\begin{pmatrix}
\frac{\pi  \left(u_1 (2 b k+b)+4 b \xi +b u_2+2 i k+2 i\right)}{2 (k+1)} \\
\frac{\pi  \left(2 i (2 i b \xi +k+1)+u_2 (2 b k+b)+b u_1\right)}{2 (k+1)}  \\
\end{pmatrix}   \right) \,, ~ \nn\\
&\T[\mathbf{(0,1)}]:  \left(
\begin{pmatrix}
\frac{k-1}{k}& -\frac{1}{k}  \\
-\frac{1}{k}  & -\frac{1}{k}    \\
\end{pmatrix} \,, ~
\begin{pmatrix}
\frac{\pi  \left(2 i (-2 i b \xi +b Q+k-2)+b (2 k-1) u_1-b u_2\right)}{2 k} \\
-\frac{\pi  \left(u_2 (2 b k+b)-4 b \xi -2 i b Q+b u_1+4 i\right)}{2 k}  \\
\end{pmatrix}    \right)
 \,, ~\nn \\
&\T[\mathbf{(0,2)} ] :   \left(
\begin{pmatrix}
1 & -1  \\
-1 & 1+k  \\
\end{pmatrix} \,, ~
\begin{pmatrix}
\pi  b \left(i Q+u_1+u_2\right)  \nn\\
\frac{1}{2} \pi  \left(-2 i b k Q-u_2 (2 b k+b)+4 b \xi -b u_1+2 i k\right)\\
\end{pmatrix}   \right)
 \,, ~\nn \\
&   \T[\mathbf{(1,0)} ]:
\left(
\begin{pmatrix}
-\frac{1}{k} & -\frac{1}{k}  \\
-\frac{1}{k} & \frac{k-1}{k} \\
\end{pmatrix} \,, ~
\begin{pmatrix}
\frac{\pi  \left(u_1 (2 b k+b)+4 b \xi -2 i b Q+b u_2+4 i\right)}{2 k}  \\
\frac{\pi  \left(2 i (2 i b \xi +b Q+k-2)+b (2 k-1) u_2-b u_1\right)}{2 k}   \\
\end{pmatrix}  \right)
 \,, ~ \nn\\
&   \T[\mathbf{(1,1)} ]:
\left(
\begin{pmatrix}
\frac{1}{1-k} & \frac{1}{k-1} \\
\frac{1}{k-1} & \frac{1}{1-k} \\
\end{pmatrix} \,, ~
\begin{pmatrix}
-\frac{\pi  b \left((2 k-1) u_1+4 \xi -u_2\right)}{2 (k-1)} \\
\frac{\pi  b \left((1-2 k) u_2+4 \xi +u_1\right)}{2 (k-1)}\\
\end{pmatrix}   \right)
\,, ~ \nn\\
&   \T[\mathbf{(1,2)} ]:
\left(
\begin{pmatrix}
0 & 1  \\
1 & k  \\
\end{pmatrix} \,, ~
\begin{pmatrix}
\pi  \left(-i b Q-b u_1-b u_2+i\right) \\
\frac{1}{2} \pi  \left(-2 i b k Q+u_2 (b-2 b k)+4 b \xi +b u_1+2 i k+2 i\right)  \\
\end{pmatrix}   \right)
\,, ~\nn \\
&   \T[\mathbf{(2,0)} ] :
\left(
\begin{pmatrix}
k+1& -1  \\
-1 & 1  \\
\end{pmatrix} \,, ~
\begin{pmatrix}
-\frac{1}{2} i \pi  \left(2 b k Q-i u_1 (2 b k+b)-4 i b \xi -i b u_2-2 k\right)  \\
\pi  b \left(i Q+u_1+u_2\right)\\
\end{pmatrix} 
\right)
\,, ~ \nn \\
&\T[\mathbf{(2,1)} ] : 
\left(
\begin{pmatrix}	
k &1  \\
1 & 0 \\
\end{pmatrix}
 \,, ~
\begin{pmatrix}
\frac{1}{2} \pi  \left(-2 i b k Q+u_1 (b-2 b k)-4 b \xi +b u_2+2 i k+2
i\right) \\
\pi  \left(-i b Q-b u_1-b u_2+i\right)\\
\end{pmatrix}    \right)	
\,.
\end{align}
Because of the exchange relation $\mathbf{n_i \leftrightarrow n_j}$, there are only two independent mirror theories with integer effective CS level matrices. We identify these theories as $\{ ~\T[\mathbf{(2,1)} ]\,,~ \T[\mathbf{(2,0)} ] ~ \}$, and they are related by the transformation $\mathbf{(0,2)}$ 
\begin{equation}
\T[\mathbf{(2,1)} ]\xlongrightarrow{ \mathbf{(0,2)}} \T[\mathbf{(2,0)}   \,.
\end{equation}
\begin{figure}
	\centering
	\includegraphics[width=1.5in]{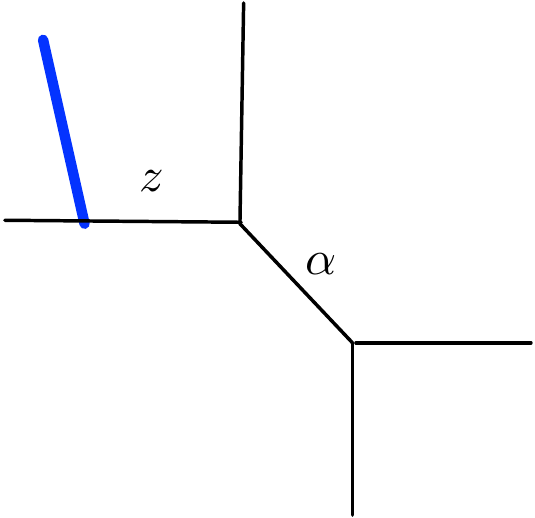} 
	\caption{The corresponding toric Calabi-Yau threefold for  theory $U(1)_k+ 1 \mathbf{C}+1  \mathbf{AC}$. }
	\label{fig:conifold}
\end{figure}
The corresponding toric diagram for $U(1)_k+ 1 \,\mathbf{C}   +1 \, \mathbf{AC} $ is shown in figure \ref{fig:conifold}. Following  \eqref{vortexNCNAC}, we get its vortex partition function
\begin{align}
\label{vortexU(1)1C1AC}
Z^{\text{vortex}} _{ U(1)_k+ 1 \,\mathbf{C}   +1 \, \mathbf{AC}} &=\sum\limits_{n=0}^{\inf} \frac{  (-\sqrt{q})^{(f+1)\, n^2}  z^n   \left(\a  ,q \right)_n}{(q,q)_n}    
\end{align}
which in combination with the one-loop part equals the vortex partition functions of $\T[\mathbf{(2,1)} ]$ and $\T[\mathbf{(2,0)} ]$ theories,
\begin{align}
Z _{ \T_{A,2}}
&=  Z^{\text{1-loop}} _{ U(1)_k+ 1 \,\mathbf{C}+1 \,\mathbf{AC}   }     \cdot Z^{\text{vortex}} _{ U(1)_k+ 1 \,\mathbf{C}+1 \,\mathbf{AC}   }   \nn\\
&= 
\sum_{ d_1,d_2=0}^{\inf}  
(-\sqrt{q})^{ \sum\limits_{i,j=1}^2 k^{eff,  \mathbf{(2,1)}}_{ij}d_i d_j} 
\frac{
	z^{d_1} \,  \a^{d_2}
} 
{ 
	(q,q)_{d_1} (q,q)_{d_2} }
\label{1c1ac1 } \\
&=\sum_{ d_1,d_2=0}^{\inf}  
(-\sqrt{q})^{ \sum\limits_{i,j=1}^2 k^{eff,  \mathbf{(2,0)}}_{ij}d_i d_j} 
\frac{
	z^{d_1} \, (\sqrt{q} \a^{-1} )^{d_2}
} 
{ 
	(q,q)_{d_1} (q,q)_{d_2} }\label{1c1ac2 }  \,,
\end{align}
where the second line is for the $\T[\mathbf{(2,1)} ]$ theory and the third line is for $\T[\mathbf{(2,0)} ]$. One can see that flipping the expansion parameter $\a \rightarrow \sqrt{q}\, \a^{-1}$ relates effective CS levels in \eqref{1c1ac1 } and \eqref{1c1ac2 },
\begin{equation}
\begin{tikzcd}[column sep=3.5pc]
k_{ij}^{eff,  \mathbf{(2,1)} } \arrow{r}{\text{flip}~ \a} & k_{ij}^{eff,  \mathbf{(2,0)} }  
\end{tikzcd}  \,.
\end{equation}
Therefore we interpret this flipping as the mirror transformation $\mathbf{(0,2)}$.

\subsection{$U(1)_k+ 3 \,\mathbf{C}$}
This theory can be turned into a particular $\T_{A,3}$ theory,
\begin{align}
Z_{S_b^3}^{(1)_{k}+ 3\mathbf{C} }
\xlongrightarrow{\mathbf{(1,1,1)}}
Z_{S_b^3}^{\T_{A,3}  }\,.	
\end{align}
Following \eqref{STofT_{1,N}}, we get the sphere partition function of the corresponding $\T_{A,3}$ theory 
\begin{align}
&Z_{S_b^3}^{\T_{A,3}  } =
\int \prod\limits_{i,j=1}^3 d y_i  \, e^{2 \pi \xi'_i y_i -i \pi k_{ij} y_i y_j } s_b\big( \frac{i Q}{2}  -y_i \big)  \,, \\
& k_{ij} =
\begin{pmatrix}
- \frac{ i (2k-1) }{6+4 k} & \frac{2 i \pi}{3+2k} &  \frac{2 i \pi}{3+2k} \\
\frac{2 i \pi}{3+2k} & - \frac{ i (2k-1) }{6+4 k}  & \frac{2 i \pi}{3+2k} \\
\frac{2 i \pi}{3+2k}  &\frac{2 i \pi}{3+2k} & - \frac{ i (2k-1) }{6+4 k} 
\end{pmatrix} \,, ~~
\xi'_i =
\begin{pmatrix}
-\frac{\pi  \left(2 k Q-4 i (k+1) u_1-8 i \xi -3 Q+2 i u_2+2 i u_3\right)}{4 k+6}
\\
-\frac{\pi  \left(2 k Q-4 i (k+1) u_2-8 i \xi -3 Q+2 i u_1+2 i u_3\right)}{4 k+6}\\
-\frac{\pi  \left(2 k Q-4 i (k+1) u_3-8 i \xi -3 Q+2 i u_1+2 i u_3\right)}{4 k+6} 
\end{pmatrix} \,.
\end{align}
By acting with mirror transformations on the sphere partition function, we get many mirror dual theories with integer effective CS level matrices $ k_{ij}^{eff, \mathbf{(n_1,n_2,n_3)}}$, 
\begin{align}
&\T[\mathbf{(0,0,2)}] :
\begin{pmatrix}
1 & 0 &1 \\
0 & 1 &1 \\
1 & 1 & k+\frac{3}{2}
\end{pmatrix}\,, &
&\T[\mathbf{(0,1,2)}] :
\begin{pmatrix}
1 & 0 &1 \\
0 & 0 &-1 \\
1 & -1 & k+\frac{1}{2}
\end{pmatrix}\,, &
&\T[\mathbf{(0,2,0)}] :
\begin{pmatrix}
1 & 1 &0 \\
1 & k+\frac{3}{2} &1 \\
0 & 1 &1 
\end{pmatrix}\,,  &   \nn\\
&\T[\mathbf{(0,2,1)}] :
\begin{pmatrix}
1 & 1 &0 \\
1 & k+\frac{1}{2} &-1 \\
0 & -1 &1 
\end{pmatrix}\,, &
&\T[\mathbf{(2,0,0)}] :
\begin{pmatrix}
k+\frac{3}{2} & 1 &1 \\
1 & 1 &0 \\
1 & 0 & 1
\end{pmatrix}\,,  &
&\T[\mathbf{(2,0,1)}] :
\begin{pmatrix}
k+\frac{1}{2}& 1 &-1 \\
1& 1 &0 \\
-1 & 0& 0
\end{pmatrix}\,,  & \nn
\\
&\T[\mathbf{(1,0,2)}] :
\left(
\begin{array}{ccc}
0 & 0 & -1 \\
0 & 1 & 1 \\
-1 & 1 & k+\frac{1}{2} \\
\end{array}
\right)\,, &
&\T[\mathbf{(1,1,2)}] :
\left(
\begin{array}{ccc}
0 & 0 & -1 \\
0 & 0 & -1 \\
-1 & -1 & k-\frac{1}{2} \\
\end{array}
\right)\,,&
&\T[\mathbf{(1,2,0)}] :
\left(
\begin{array}{ccc}
0 & -1 & 0 \\
-1 & k+\frac{1}{2} & 1 \\
0 & 1 & 1 \\
\end{array}
\right)\,,     &  \nn\\
&\T[\mathbf{(1,2,1)}] :
\left(
\begin{array}{ccc}
0 & -1 & 0 \\
-1 & k-\frac{1}{2} & -1 \\
0 & -1 & 0 \\
\end{array}
\right)\,,&
&\T[\mathbf{(2,1,0)}] :
\left(
\begin{array}{ccc}
k+\frac{1}{2} & -1 & 1 \\
-1 & 0 & 0 \\
1 & 0 & 1 \\
\end{array}
\right)\,, &
&\T[\mathbf{(2,1,1)}] :
\left(
\begin{array}{ccc}
k-\frac{1}{2} & -1 & -1 \\
-1 & 0 & 0 \\
-1 & 0 & 0 \\
\end{array}
\right) \,.	&
 \end{align}
Because of the exchange relation $\mathbf{n_i \leftrightarrow n_j}$, there are only four independent theories. We choose them to be $\{  \T[\mathbf{(2,0,0)} ]\,, \T[\mathbf{(2,0,1)} ]  \,,\T[\mathbf{(2,1,0)} ] \,, \T[\mathbf{(2,1,1)} ] \}$, and their effective CS level matrices and effective FI parameters are as follows:
\begin{align}
\T[\mathbf{(2,0,0)} ] &:~
\left(
\left(
\begin{array}{ccc}
k+\frac{3}{2} & 1 & 1 \\
1 & 1 & 0 \\
1 & 0 & 1 \\
\end{array}
\right) \,,~
\left( 
\begin{array}{c}
\frac{1}{2} \pi  \left(-2 i b k Q-2 b (k+1) u_1-4 b \xi -4 i b Q+ b u_2+b u_3+2 i k+7 i\right)
\nn\\
\pi  \left(-i b Q-b u_1+b u_2+2 i\right)   \nn\\
\pi  \left(-i b Q-b u_1+b u_3+2 
i\right)  \nn\\
\end{array}
\right)    \right)    
  \,,  \\
\T[\mathbf{(2,0,1)} ] &:~
\left( \left(
\begin{array}{ccc}
k+\frac{1}{2} & 1 & -1 \\
1 & 1 & 0 \\
-1 & 0 & 0 \\
\end{array}
\right) \,,~
\left( 
\begin{array}{c}
\frac{1}{2} \pi  \left(-2 i b k Q-2 b k u_1-4 b \xi +b u_2-b u_3+2 i k+i\right)
\nn\\
\pi  \left(-i b Q-b u_1+b u_2+2 i\right)
\nn\\
\pi  \left(i b Q+b u_1-b u_3-i\right) 
\nn\\
\end{array}
\right) \right)  \,,  \\
\T[\mathbf{(2,1,0)} ] &:~
\left(  \left(
\begin{array}{ccc}
k+\frac{1}{2} & -1 & 1 \\
-1 & 0 & 0 \\
1 & 0 & 1 \\
\end{array}
\right) \,,~
\left( 
\begin{array}{c}
\frac{1}{2} \pi  \left(-2 i b k Q-2 b k u_1-4 b \xi -b u_2+b u_3+2 i k+i\right)
\nn\\
\pi  \left(i b Q+b u_1-b u_2-i\right)
\nn\\
\pi  \left(-i b Q-b u_1+b u_3+2 i\right)
\nn\\
\end{array}
\right)  \right)  \,,  \\
\T[\mathbf{(2,1,1)} ] &:~
\left(\left(
\begin{array}{ccc}
k-\frac{1}{2} & -1 & -1 \\
-1 & 0 & 0 \\
-1 & 0 & 0 \\
\end{array}
\right)  \,,~
\left(
\begin{array}{c}
\frac{1}{2} \pi  \left(-2 i b k Q-2 b (k-1) u_1-4 b \xi +4 i b Q-b u_2-b u_3+2 i k-5
i\right) \\
\pi  \left(i b Q+b u_1-b u_2-i\right) \\
\pi  \left(i b Q+b u_1-b u_3-i\right) \\
\end{array}
\right)  \right)
  \,.
\end{align}
These four mirror dual theories are related by
\begin{equation}\label{3CTmirror}
\begin{tikzcd}	[column sep=3.5pc]
\T[\mathbf{(2,0,0)}]  \arrow{d}{ \mathbf{ (0,0,1)} } \arrow{r}{  \mathbf{ (0,1,0)} } &  \T[\mathbf{(2,1,0)} ] \arrow{d}{  \mathbf{ (0,0,1)}  } \\
\T[ \mathbf{(2,0,1)}]    \arrow{r}{ \mathbf{ (0,1,0)} } &
\T[ \mathbf{(2,1,1)} ] \,\,.
\end{tikzcd} 
\end{equation}
\begin{figure}
	\centering
	\includegraphics[width=1.5in]{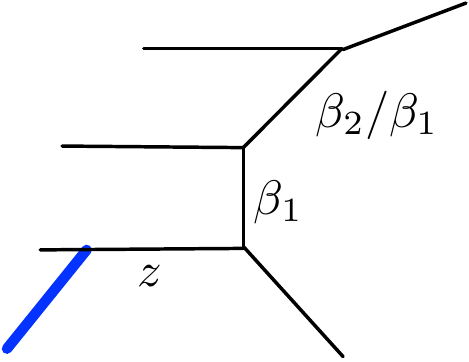} 
	\caption{The corresponding toric Calabi-Yau threefold for theory $U(1)_k+ 3 \,\mathbf{C}$. Note that putting the open topological brane (marked in blue) on various horizontal lines gives rise to the same theory.}
	\label{fig:(1)_k+3C}
\end{figure}
The toric diagram for $U(1)_k+ 3 \,\mathbf{C}$ is shown in figure \ref{fig:(1)_k+3C}.
Using \eqref{vortexNCNAC}, its vortex partition function is given by
\begin{align}
\label{vortexU(1)3C}
Z^{\text{vortex}} _{ U(1)_k+ 3 \,\mathbf{C}} &=\sum\limits_{n=0}^{\inf} \frac{  (-\sqrt{q})^{(f+1)\, n^2}  z^n}{(q,q)_n} 
\frac{1}
{ 	\left(\b_1 ,q \right)_n \left(\b_2  ,q \right)_n
}   \,,
\end{align}
which can be written in terms of vortex partitions of the above four dual theories:
\begin{align}
Z_{\T_{A,3}} &=	Z^{\text{1-loop}} _{ U(1)_k+ 3 \,\mathbf{C}  } \cdot Z^{\text{vortex}} _{ U(1)_k+ 3 \,\mathbf{C}  }   \\
&=\sum_{ d_1,d_2,d_3=0}^{\inf}  
(-\sqrt{q})^{ \sum\limits_{i,j=1}^3 k^{eff, \mathbf{(2,0,0)}}_{ij}d_i d_j} 
\frac{
z^{d_1}(\b_1/\sqrt{q})^{d_2}(\b_2/\sqrt{q})^{d_3}
} 
{ 
	(q,q)_{d_1}(q,q)_{d_2}(q,q)_{d_3} }  \\
&=\sum_{ d_1,d_2,d_3=0}^{\inf}  
(-\sqrt{q})^{ \sum\limits_{i,j=1}^3 k^{eff, \mathbf{(2,1,0)}}_{ij}d_i d_j} 
\frac{
	z^{d_1}(q\, \b_1^{-1})^{d_2}(\b_2/\sqrt{q})^{d_3}
} 
{ 
	(q,q)_{d_1}(q,q)_{d_2}(q,q)_{d_3} }  \\
&=\sum_{ d_1,d_2,d_3=0}^{\inf}  
(-\sqrt{q})^{ \sum\limits_{i,j=1}^3 k^{eff, \mathbf{(2,0,1)}}_{ij}d_i d_j} 
\frac{
	z^{d_1}(\b_1/\sqrt{q})^{d_2}(q\,\b_2^{-1})^{d_3}
} 
{ 
	(q,q)_{d_1}(q,q)_{d_2}(q,q)_{d_3} }   \\
&=\sum_{ d_1,d_2,d_3=0}^{\inf}  
(-\sqrt{q})^{ \sum\limits_{i,j=1}^3 k^{eff, \mathbf{(2,1,1)}}_{ij}d_i d_j} 
\frac{
	z^{d_1}(q \,\b_1^{-1})^{d_2}(q\,\b_2^{-1})^{d_3}
} 
{ 
	(q,q)_{d_1}(q,q)_{d_2}(q,q)_{d_3} }   \,.
\end{align}
It is obvious that mixed CS level matrices for these mirror dual theories are related by flipping closed K\"ahler parameters $\b_i$,
\begin{equation}
\begin{tikzcd}[column sep=3.5pc]
k_{ij}^{eff,  \mathbf{(2,0,0)} } \arrow{d}{ \text{flip}~ \b_2} \arrow{r}{\text{flip}~ \b_1} & k_{ij}^{eff,  \mathbf{(2,1,0)} } \arrow{d}{ \text{flip}~ \b_2 } \\
k_{ij}^{eff,  \mathbf{(2,0,1)} }    \arrow{r}{\text{flip}~ \b_1 } &
k_{ij}^{eff,  \mathbf{(2,1,1)} }   \,.
\end{tikzcd}
\end{equation}
Therefore, to match with \eqref{3CTmirror}, the flipping $\b_1$ should correspond to mirror transformation $\mathbf{(0,1,0 )}$ and flipping $\b_2$ corresponds to $\mathbf{(0,0,1 )}$.
This confirms the fact that mirror transformations are interpreted as flipping K\"ahler parameter $x_i$ of vortex partition functions of $\T_{A, N}$ theories corresponding to strip Calabi-Yau threefolds.

\subsubsection*{Tong's mirror pair}\label{tongpair}
When $k=-3/2$, the dual $\T_{A,3}$ theory given by  \eqref{STofT_{1,N}}  is problematic because of poles in $\tilde{k}_{ij}$. 
%At first sight, $U(1)_{-3/2}+ 3 \,\mathbf{C}$ cannot be transformed into particular $\T_{A,3}$ theories. 
Nevertheless, it is possible to bypass this pole in $\tilde{k}_{ij}$ and still get a well-defined $\T_{A,3}$ theories. The procedure of addressing this problem is as follows: first we do not give value to $k$ and continue acting  $\mathbf{(2,0,0)}$ on the partition function,  and at the end we set $k=-3/2$. This leads to a well defined partition function that can be viewed as the original theory as well. This new original $\T'_{A,3}$ theory is given by mirror transformation $\mathbf{(1,1,1)}+ \mathbf{(2,0,0)} =\mathbf{(0,1,1)} $. More explicitly, its sphere partition function is obtained in two steps
\begin{align}\label{bypass}
&Z_{S_b^3}^{U(1)_{-3/2}+3 \mathbf{C}} \xlongrightarrow
{\mathbf{(1,1,1)}} \bullet \xlongrightarrow
{\mathbf{(2,0,0)}} Z^{\T'_{A,3}}_{S_b^3}  	
\end{align}
where
\begin{align}
Z^{\T'_{A,3}}_{S_b^3} &=
\int dx_1 dx_2 dx_3\,e^{ \textbf{CS term} }
s_b\big(\frac{i Q}{2}-x_1\big) s_b\big(\frac{i Q}{2}-x_2\big)
s_b\big(\frac{i Q}{2}-x_3\big)  \,,\\
\textbf{CS term}&= \frac{1  }{2} \pi i\,(x_1^2-x_2^2-x_3^2 ) - \pi \left( \frac{Q}{2} +i u_1 -i u_2  \right) x_2 -\pi \left( \frac{Q}{2}  +i u_1-i u_3   \right)   \nn\\
	&\qquad\qquad\quad -\pi \left( Q+ 2 i \xi -\frac{i}{2} (u_1+u_2+u_3) \right)  -2 \pi i\, (x_2+x_3) x_1  \,.
\end{align}
Furthermore, when acting with the mirror transformation $\mathbf{(1,0,0)}$ on this new original theory $\T'_{A,3}$, one gets $\T'_{A,3}[ \mathbf{(1,0,0)}]$,
\begin{align} 
&Z_{S_b^3}^{U(1)_{-3/2}+3 \mathbf{C}} \xrightarrow
{\mathbf{(1,1,1)}} \bullet \xrightarrow
{\mathbf{(2,0,0)} } \bullet \xrightarrow{ \mathbf{(1,0,0)}} Z_{S_b^3}^{\T'_{A,3}[ \mathbf{(1,0,0)}]}   \,.
\end{align}
Here, we encounter quiver reduction for the theory $\T'_{A,3}[ \mathbf{(1,0,0)}]$ that turns out to have a reduced quiver. Its sphere partition function, after shifting parameters $x_2 \rightarrow -x_2, ~u_2 \rightarrow -3 i\,Q +4\xi-u_1-u_3$, is the following:
%has the following sphere partition function 
\begin{align}
Z_{S_b^3}^{\T'_{A,3}[ \mathbf{(1,0,0)}]}   & =	\int dx_2 dx_3\,
e^{\textbf{CS terms}    }  
 s_b\big( \frac{i Q}{2} + x_2)
s_b\big( \frac{i Q}{2} - x_3)
s_b\big( \frac{i Q}{2} - x_2+x_3) \,,  \label{bypass1}  \\
 \textbf{CS terms}& =-i \pi ( x_2^2+x_3^2- x_2 x_3   )  -i \pi (u_1-u_3) x_3 -\pi (3 Q+4 i\,\xi -2i u_2 -i u_3 )x_2   \,.\nn
\end{align}
The integral dimension for this theory is two, and hence the gauge group is $U(1)\times U(1)$.
Since 
$
\mathbf{(2,0,0)}+\mathbf{(1,0,0) }=\mathbf{(0,0,0)}  
$,
\eqref{bypass1} is equivalent to the problematic sphere partition function given in \eqref{STofT_{1,N}} with $k=-3/2$.
The associated
bare CS level matrix for \eqref{bypass1}  is
\begin{align}
k_{ij} = 
\left(
\begin{array}{cc}
1  & -\frac{1}{2}   \\
-\frac{1}{2}  & 1
\end{array}
\right) \,,
\end{align}
and the associated chiral multiplets have charges $(-1,0),~ (1,-1),~(0,1)$, respectively. 
It is easy to draw its quiver 
\begin{align}
	[1]- U(1)-U(1)-[1]   \,.
	\end{align}
 Interestingly, we obtain the mirror pair found by Dorey and Tong in \cite{Dorey:1999rb},
\begin{align}\label{Tongdual}
\begin{array}{c}
U(1)_{-3/2}-[3] \,,\\
\text{with}~ k=-3/2\,, 
\\
\text{and}~ ~k^{eff}=0
\end{array}
\quad\xleftrightarrow{\mathbf{~(1,1,1)~}} \quad 
\begin{array}{c}
[1]- U(1)-U(1)-[1] 
\quad\text{with}\quad
k_{ij} = 
\left(
\begin{array}{cc}
1  & -\frac{1}{2}   \\
-\frac{1}{2}  & 1
\end{array}
\right) \,.\\
\text{and}~~
k^{eff}_{ij} =
\left(
\begin{array}{cc}
2  & -1   \\
-1  & 2
\end{array}
\right)
\end{array}
\end{align}
In this case the mirror transformation is $\mathbf{(1,1,1)}$.
This example illustrates the fact that mirror transformations can be used to verify and derive dualities with the help of $\T_{A,N}$ theories.

\subsection{$U(1)_k+ 2 \, \mathbf{C}+1 \, \mathbf{AC}$}
The sphere partition function for this theory is
\begin{align}
&Z_{S_b^3}^{U(1)_k+ 2 \, \mathbf{C}+1 \, \mathbf{AC}}  = \int d x\, e^{2 \pi \xi x -i \pi k x^2 } s_b\big( \frac{ i Q}{2} +x +\frac{u_1}{2}   \big)
s_b\big( \frac{ i Q}{2} -x +\frac{u_2}{2}   \big)
s_b\big( \frac{ i Q}{2} +x +\frac{u_3}{2}   \big)  \,.
\end{align}
Because of parity anomaly, the bare CS level $k \in \mathbb{Z}+1/2$.
Mirror transformation $\mathbf{(1,1,1)}$ turns this theory into type $\T_{A,3}$,
\begin{align}
Z_{S_b^3}^{(1)_{k}+2\mathbf{C}+1\mathbf{AC} }
\xlongrightarrow{\mathbf{(1,1,1)}}
Z_{S_b^3}^{\T_{A,3}  }\,,
\end{align}
where the open partition function of $\T_{A,3}$ in this case is
\begin{align}
 &Z_{S_b^3}^{\T_{A,3}  } =
\int \prod\limits_{i,j=1}^3 d y_i  \, e^{2 \pi \xi'_i y_i -i \pi k_{ij} y_i y_j } s_b\big( \frac{i Q}{2}  -y_i \big)  \,, \\
& k_{ij} =
\begin{pmatrix}
- \frac{ i (2k-1) }{6+4 k} &- \frac{2 i \pi}{3+2k} &  \frac{2 i \pi}{3+2k} \\
- \frac{2 i \pi}{3+2k} & - \frac{ i (2k-1) }{6+4 k}  & -\frac{2 i \pi}{3+2k} \\
\frac{2 i \pi}{3+2k}  &-\frac{2 i \pi}{3+2k} & - \frac{ i (2k-1) }{6+4 k} 
\end{pmatrix} \,, ~~
\xi'_i =
\begin{pmatrix}
-\frac{  \pi (  (1+2k)Q -8 i \xi-4i (1+k)u_1 -2 iu_2 +2i u_3    ) }{6+4k}\\
-\frac{  i \pi (  -i(5+2k)Q +8\xi -2u_1 -4(1+k)u_2-2 u_3   )         }{6+4k}  \\
-\frac{ i \pi( -i(1+2k)Q -8 \xi+2u_1 -2 u_2 -4(1+k) u_3      )   }{ 6+4k } 
\end{pmatrix}  \,.
\end{align}
Similarly as before, we list all integer effective CS level matrices obtained by mirror transformations	
\begin{align}
&\T[\mathbf{(0,0,2) }]:
\left(
\begin{array}{ccc}
1 & 0 & 1 \\
0 & 1 & -1 \\
1 & -1 & k+\frac{3}{2} \\
\end{array}
\right)\,,~~
\T[\mathbf{(0,1,2) }]:
\left(
\begin{array}{ccc}
1 & 0 & 1 \\
0 & 0 & 1 \\
1 & 1 & k+\frac{1}{2} \\
\end{array}
\right)\,,\qquad
\T[\mathbf{(0,2,0) }]:
\left(
\begin{array}{ccc}
1 & -1 & 0 \\
-1 & k+\frac{3}{2} & -1 \\
0 & -1 & 1 \\
\end{array}
\right) \,,    \nn\\
& \T[\mathbf{(0,2,1) }]:
\left(
\begin{array}{ccc}
1 & -1 & 0 \\
-1 & k+\frac{1}{2} & 1 \\
0 & 1 & 0 \\
\end{array}
\right)\,,~~
\T[\mathbf{(1,0,2) }]:
\left(
\begin{array}{ccc}
0 & 0 & -1 \\
0 & 1 & -1 \\
-1 & -1 & k+\frac{1}{2} \\
\end{array}
\right)\,,~
\T[\mathbf{(1,1,2) }]:
\left(
\begin{array}{ccc}
0 & 0 & -1 \\
0 & 0 & 1 \\
-1 & 1 & k-\frac{1}{2} \\
\end{array}
\right)\,,     \nn\\
&\T[\mathbf{(1,2,0) }]:
\left(
\begin{array}{ccc}
0 & 1 & 0 \\
1 & k+\frac{1}{2} & -1 \\
0 & -1 & 1 \\
\end{array}
\right)\,,~~
\T[\mathbf{(1,2,1) }]:
\left(
\begin{array}{ccc}
0 & 1 & 0 \\
1 & k-\frac{1}{2} & 1 \\
0 & 1 & 0 \\
\end{array}
\right)\,,\quad~~
\T[\mathbf{(2,0,0) }]:
\left(
\begin{array}{ccc}
k+\frac{3}{2} & -1 & 1 \\
-1 & 1 & 0 \\
1 & 0 & 1 \\
\end{array}
\right)\,,           \nn \\
&\T[\mathbf{(2,0,1) }]:
\left(
\begin{array}{ccc}
k+\frac{1}{2} & -1 & -1 \\
-1 & 1 & 0 \\
-1 & 0 & 0 \\
\end{array}
\right)\,,~
\T[\mathbf{(2,1,0) }]:
\left(
\begin{array}{ccc}
k+\frac{1}{2} & 1 & 1 \\
1 & 0 & 0 \\
1 & 0 & 1 \\
\end{array}
\right)\,,\quad
\T[\mathbf{(2,1,1) }]:
\left(
\begin{array}{ccc}
k-\frac{1}{2} & 1 & -1 \\
1 & 0 & 0 \\
-1 & 0 & 0 \\
\end{array}
\right)\,,
\end{align}
which satisfy exchange equivalence $\mathbf{n_i \leftrightarrow n_j}$, so there are only four independent theories that we choose them to be \begin{align}  \label{2c1acclass}
\{\,T[\mathbf{(2,0,0)} ], \, \T[\mathbf{(2,0,1)} ],\, 
\T[\mathbf{(2,1,0)} ], \,   \T[\mathbf{(2,1,1)} ] \, \}   \,.
\end{align}
The associated effective CS levels and effective FI parameters are as follows
\begin{align}\label{(1)+k+2C+1ACmirror}
&\T[\mathbf{(2,1,0)} ]: \left(
\left(\begin{array}{ccc}
\frac{1}{2} +k  & 1& 1 \\
1&0& 0\\
1 &0  & 1
\end{array}  \right)\,, ~
\begin{pmatrix}
\frac{1}{2} \pi  \left(-2 i b k Q-2 b k u_1-4 b \xi -2 i b Q+b u_2+b u_3+2 i k+5 i\right)   \\
\pi  \left(-i b Q-b u_1-b u_2+i\right)   \\
\pi  \left(-i b Q-b u_1+b u_3+2 i\right)
\end{pmatrix}  
\right) \,, \\
&\T[\mathbf{(2,0,0)}]:   \left(
\left(
\begin{array}{ccc}
k+\frac{3}{2} & -1 & 1 \\
-1 & 1 & 0 \\
1 & 0 & 1 \\
\end{array}
\right)  \,, ~
\left(
\begin{array}{c}
\frac{1}{2} \pi  \left(-2 i b k Q-2 b (k+1) u_1-4 b \xi -2 i b Q-b u_2+b u_3+2 i k+3
i\right) \\
\pi  b \left(i Q+u_1+u_2\right) \\
\pi  \left(-i b Q-b u_1+b u_3+2 i\right) \\
\end{array}
\right)   \right)  \,,\\
&  \T[\mathbf{(2,0,1)}]:   \left( 
\left(
\begin{array}{ccc}
k+\frac{1}{2} & -1 & -1 \\
-1 & 1 & 0 \\
-1 & 0 & 0 \\
\end{array}
\right)  \,,~
\left(
\begin{array}{c}
-\frac{1}{2} i \pi  \left(2 b k Q-2 i b k u_1-4 i b \xi -2 b Q-i b u_2-i b u_3-2
k+3\right) \\
\pi  b \left(i Q+u_1+u_2\right) \\
\pi  \left(i b Q+b u_1-b u_3-i\right) \\
\end{array}
\right) \right) \,,\\
&\T[\mathbf{(2,1,1)}]:  \left(
\left(
\begin{array}{ccc}
k-\frac{1}{2} & 1 & -1 \\
1 & 0 & 0 \\
-1 & 0 & 0 \\
\end{array}
\right) \,,~\left(
\begin{array}{c}
\frac{1}{2} \pi  \left(-2 i b k Q-2 b (k-1) u_1-4 b \xi +2 i b Q+b u_2-b u_3+2 i
k-i\right) \\
\pi  \left(-i b Q-b u_1-b u_2+i\right) \\
\pi  \left(i b Q+b u_1-b u_3-i\right) \\
\end{array}
\right)   \right)  \,.
\end{align}
These four mirror dual theories are related by mirror transformations
\begin{equation}
\begin{tikzcd}[column sep=3.5pc]
\T[\mathbf{(2,1,0)}]  \arrow{d}{ \mathbf{ (0,0,1)} } \arrow{r}{  \mathbf{ (0,2,0)} } &  \T[\mathbf{(2,0,0)} ] \arrow{d}{  \mathbf{ (0,0,1)}  } \\
\T[ \mathbf{(2,1,1)}]    \arrow{r}{ \mathbf{ (0,2,0)} } &
\T[ \mathbf{(2,0,1)} ]  \,.
\end{tikzcd} 
\end{equation}
\begin{figure}
	\centering
	\includegraphics[width=1.6in]{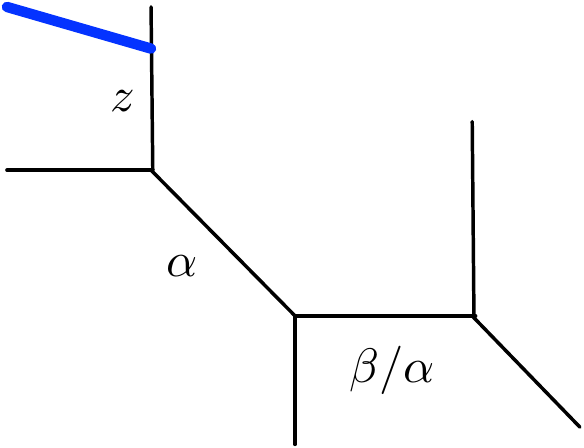} 
	\caption{The corresponding toric Calabi-Yau threefold for theory $U(1)_k+ 2 \mathbf{C} + 1 \mathbf{AC}$. 
		Note that the vortex partition function is invariant under the flop transition on closed K\"ahler parameter $\a$.
	}
	\label{fig:(1)_k+2C+1AC}
\end{figure}
The  toric diagram for this example is shown in figure \ref{fig:(1)_k+2C+1AC}. 
The corresponding vortex partition function is 
\begin{align}
\label{vortexU(1)2C1AC}
Z^{\text{vortex}} _{ U(1)_k+ 2 \,\mathbf{C}   +1 \, \mathbf{AC}} &=\sum\limits_{n=0}^{\inf} \frac{  (-\sqrt{q})^{(f+1)\, n^2}  z^n}{(q,q)_n} 
\frac{\left(\a  ,q \right)_n}
{ 	\left(\b,q \right)_n 
}    \,,
\end{align}
which along with the one-loop part is equivalent to the vortex partition functions of four mirror dual theories mentioned in \eqref{2c1acclass},
\begin{align}
Z_{\T_{A,3}} &=	Z^{\text{1-loop}} _{ U(1)_k+ 2 \,\mathbf{C}   +1 \, \mathbf{AC}} \cdot Z^{\text{vortex}} _{ U(1)_k+ 2 \,\mathbf{C}   +1 \, \mathbf{AC}}   \\
&=\sum_{ d_1,d_2,d_3=0}^{\inf}  
(-\sqrt{q})^{ \sum\limits_{i,j=1}^3 k^{eff, \mathbf{(2,1,0)}}_{ij}d_i d_j} 
\frac{
	z^{d_1}   \a^{d_2}  ( \b/\sqrt{q} )^{d_3}
} 
{ 
	(q,q)_{d_1}(q,q)_{d_2}(q,q)_{d_3} }  \\
&=\sum_{ d_1,d_2,d_3=0}^{\inf}  
(-\sqrt{q})^{ \sum\limits_{i,j=1}^3 k^{eff, \mathbf{(2,0,0)}}_{ij}d_i d_j} 
\frac{
	z^{d_1}   ( \sqrt{q} \a^{-1} )^{d_2}  ( \b/\sqrt{q} )^{d_3}
} 
{ 
	(q,q)_{d_1}(q,q)_{d_2}(q,q)_{d_3} }  \\
&=\sum_{ d_1,d_2,d_3=0}^{\inf}  
(-\sqrt{q})^{ \sum\limits_{i,j=1}^3 k^{eff, \mathbf{(2,0,1)}}_{ij}d_i d_j} 
\frac{
	z^{d_1}   (\sqrt{q} \a^{-1} )^{d_2}  ( q\, \b^{-1} )^{d_3}
} 
{ 
	(q,q)_{d_1}(q,q)_{d_2}(q,q)_{d_3} }  \\
&=\sum_{ d_1,d_2,d_3=0}^{\inf}  
(-\sqrt{q})^{ \sum\limits_{i,j=1}^3 k^{eff, \mathbf{(2,1,1)}}_{ij}d_i d_j} 
\frac{
	z^{d_1}   \a^{d_2}  ( q \, \b^{-1})^{d_3}
} 
{ 
	(q,q)_{d_1}(q,q)_{d_2}(q,q)_{d_3} }    \,.	
\end{align}
It is obvious that flipping $\a \rightarrow \sqrt{q} \a^{-1}$ and $ \b \rightarrow \sqrt{q} \b^{-1}$ relates their effective CS level matrices
\begin{equation}
\begin{tikzcd}[column sep=3.5pc]
k_{ij}^{eff,  \mathbf{(2,1,0)} } \arrow{d}{ \text{flip}~ \b} \arrow{r}{\text{flip}~ \a} & k_{ij}^{eff,  \mathbf{(2,0,0)} } \arrow{d}{ \text{flip}~ \b} \\
k_{ij}^{eff,  \mathbf{(2,1,1)} }    \arrow{r}{\text{flip}~ \a } &
k_{ij}^{eff,  \mathbf{(2,0,1)} }    \,.
\end{tikzcd}  
\end{equation}
Once again, this confirms that mirror symmetry can be interpreted as flipping closed K\"ahler parameters  in vortex partition functions.

\subsection{$ [1]-U(1)_{k_1}-U(1)_{k_2}-[1] $}
This quiver theory has three chiral multiplets with charges $(1,0), (p_1,p_2), (0,1)$, respectively. The associated sphere partition function is given by
\begin{align}
Z_{S_b^3}^{[1]-(1)_{k_1}-(1)_{k_2}-[1]}&= \int  dx_1 dx_2 \, e^{ - i k_1 \pi x_1^2 -i k_2 \pi x_2^2 + 2 \pi i (\xi_1 x_1 +\xi_2 x_2  )}  \nn\\
&\times s_b \big(\frac{iQ}{2} +x_1+\frac{u_1}{2} )  s_b \big(\frac{iQ}{2} +x_2+\frac{u_2}{2} )  s_b \big(\frac{iQ}{2} +p_1 x_1+p_2 x_2+\frac{u_1}{2} ) \,.\end{align}
After redefining parameters 
\begin{align}
u_1 :=\frac{\log Y_1}{b \pi}, ~ u_2 := \frac{\log Y_2}{b \pi} ,~u_3:= \frac{\log(-q^{(p_1+p_2-1)/2}\, Y_3  -i \pi( p_1+p_2))  }{b \pi} \,,
\end{align}
we get the associated effective superpotential in the semiclassical limit
\begin{align}
\Weff_{[1]-(1)_{k_1}-(1)_{k_2}-[1] } =\,& \Li_2(X_1Y_1)+\Li_2(X_2 Y_2)+\Li_2(X_1^{p_1}X_2^{p_2}Y_3 ) \nn\\
&+\frac{1}{2} \big( k_1 +\frac{1+p_1^2}{2} \big) \log X_1^2 +
\frac{1}{2} \big( k_2 +\frac{1+p_2^2}{2} \big) \log X_2^2 +
\frac{p_1 p_2}{2} \log X_1 \log X_2  \nn\\
&+\sum\limits_{l=1}^2 \big((1+p_l)  \pi i+ \log Y_1+p_l \log Y_3+ 2 \pi i \, k_1 -k_l \log q -4 b \pi \xi_l \big) \log X_l   \,.\nn\\
\end{align}
The associated effective CS level matrix is
\begin{align}
k_{ij}^{eff}= 
\left(
\begin{array}{cc}
k_1 +\frac{1+p_1^2}{2} &\frac{p_1 p_2}{2}\\
\frac{p_1 p_2}{2}& k_2 +\frac{1+p_2^2}{2}
\end{array}
\right) \,.
\end{align}
Similarly as before, mirror transformation $\mathbf{(1,1,1)}$ turns this quiver theory into some particular $\T_{A,3}$ theories,
\begin{align}
Z_{S_b^3}^{[1]-(1)_{k_1}-(1)_{k_2}-[1]}
\xlongrightarrow{\mathbf{(1,1,1)} }
Z_{S_b^3}^{ \T_{A,3}} \,.
\end{align}
We list some effective CS level matrices given by mirror transformations
\begin{align}
& \T[\mathbf{(0,2,2)}]:
\left(
\begin{array}{ccc}
1 & \frac{1}{p_1} & -\frac{p_2}{p_1} \\
\frac{1}{p_1} & \frac{2 k_1+p_1^2+1}{2 p_1^2} & -\frac{2 k_1 p_2+p_2}{2 p_1^2} \\
-\frac{p_2}{p_1} & -\frac{2 k_1 p_2+p_2}{2 p_1^2} & \frac{k_1
	p_2^2}{p_1^2}+k_2+\frac{1}{2} \left(\frac{p_2^2}{p_1^2}+1\right) \\
\end{array}
\right) \,,\\
&\T[ \mathbf{(1,2,2)} ]:
\left(
\begin{array}{ccc}
0 & -\frac{1}{p_1} & \frac{p_2}{p_1} \\
-\frac{1}{p_1} & \frac{2 k_1+p_1^2-1}{2 p_1^2} & \frac{p_2-2 k_1 p_2}{2 p_1^2} \\
\frac{p_2}{p_1} & \frac{p_2-2 k_1 p_2}{2 p_1^2} & \frac{k_1
	p_2^2}{p_1^2}+k_2-\frac{p_2^2}{2 p_1^2}+\frac{1}{2} \\
\end{array}
\right)  \,,\\
&\T[\mathbf{(2,0,2)}]:
\left(
\begin{array}{ccc}
\frac{1}{2} \left(2 k_1+p_1^2+1\right) & p_1 & \frac{p_1 p_2}{2}
\\
p_2 & 1 & p_2\\
\frac{p_1 p_2}{2} & p_2 & \frac{1}{2} \left(2 k_2+p_2^2+1\right)
\\
\end{array}
\right) \,,       \\
&\T[\mathbf{(2,1,2)}]:
\left(
\begin{array}{ccc}
k_1-\frac{p_1^2}{2}+\frac{1}{2} & -p_1 & -\frac{1}{2} p_1 p_2 \\
-p_1 & 0 & -p_2 \\
-\frac{1}{2} p_1 p_2 & -p_2 & k_2-\frac{p_2^2}{2}+\frac{1}{2} \\
\end{array}
\right) \,,\\
& \T[\mathbf{(2,2,0)}]:
\left(
\begin{array}{ccc}
\frac{2 k_2 p_1^2+p_1^2+p_2^2}{2 p_2^2}+k_1 & -\frac{2 k_2 p_1+p_1}{2 p_2^2} &
-\frac{p_1}{p_2} \\
-\frac{2 k_2 p_1+p_1}{2 p_2^2} & \frac{2 k_2+p_2^2+1}{2 p_2^2} & \frac{1}{p_2} \\
-\frac{p_1}{p_2} & \frac{1}{p_2} & 1 \\
\end{array}
\right)\,,\\
&\T[\mathbf{(2,2,1)} ]:
\left(
\begin{array}{ccc}
\frac{2 k_2 p_1^2-p_1^2+p_2^2}{2 p_2^2}+k_1 & \frac{p_1-2 k_2 p_1}{2 p_2^2} &
\frac{p_1}{p_2} \\
\frac{p_1-2 k_2 p_1}{2 p_2^2} & \frac{2 k_2+p_2^2-1}{2 p_2^2} & -\frac{1}{p_2} \\
\frac{p_1}{p_2} & -\frac{1}{p_2} & 0 \\
\end{array}
\right) \,.
\end{align}
It is obvious that if charges $p_1$ and $p_2$ for the bifundamental multiplet are chosen properly,  there could be many anomaly free mirror dual theories with integer effective CS levels.

\section{ Knot polynomials }\label{section-4}

Mirror symmetry is also important in knot theory, because many knot invariants can be engineered by gauge theories.
The theories $U(1)_k+ N_C \mathbf{C}+N_{AC} \mathbf{AC}$ discussed in section 3 actually correspond to the unknot. However, in this work we expect that mirror transformations could be applied to generic knots.

In \cite{Kucharski:2017poe,Kucharski:2017ogk}, it is found that the HOMFLY-PT polynomials of various knots can be lifted to the form
\begin{align}\label{quivergen}
P^{K}(a,x,q) \xrightarrow{\text{lift}} P^{Q_K}(\textbf{x},q) :=\sum_{d_1,...,d_N=0}^{\inf}  
(-\sqrt{q})^{ \sum\limits_{i,j=1}^N C_{ij}d_i d_j} 
\frac{ 
	x_1^{d_1}\dots x_N^{d_N}
} 
{ 
	(q,q)_{d_1}\dots (q,q)_{d_N} }
\,,
\end{align}
which implies that different knots correspond to matrices $C_{ij}$. This relation is called the knots-quivers correspondence (KQ) in \cite{Kucharski:2017ogk}. \footnote{In \cite{Kucharski:2017ogk,Ekholm:2018eee}, $C_{ij}$ is called \textit{quiver} following the notation in quiver representation theory.} Moreover, some identifications need to be imposed on variables $x_i$,
\begin{align}\label{KQxirelation}
x_i=x \,a^{a_i} q^{\frac{q_i-C_{ii}}{2}}  (-\textbf{t})^{\frac{C_{ii}}{2}}   
\end{align}
in order to ensure that
\begin{align}
P^K(a,x,q) = P^{Q_K}\big(x_i=x \,a^{a_i} q^{\frac{q_i-C_{ii}}{2}}  (-\textbf{t})^{\frac{C_{ii}}{2}}     ,q  \big)
\end{align}
where parameter $-\textbf{t}=1$ in the unrefined limit $q=t$.
On the other hand, 3d/3d correspondence claims that colored HOMFLY-PT polynomials are equal to vortex partition functions of certain 3d $\N=2$ theories \cite{Dimofte:2010tz,Dimofte:2011ju}. 
Inspired by this argument and the form \eqref{quivergen}, it is conjectured in \cite{Ekholm:2018eee} that the lifted version $P^{Q_K}(\textbf{x},t)$ also corresponds to certain 3d $\N=2$ theories $T[Q_K]$ whose vortex partition functions in the semiclassical limit take form
\begin{align}
&P^{Q_K}(\textbf{x},q)\xrightarrow{\hbar\rightarrow 0} \int \prod\limits_i \frac{d y_i}{y_i} \exp \frac{1}{  \hbar   }\Big(  \widetilde\W_{T[Q_K] }  ( \mathbf{x} ,\mathbf{y}  )  + O(\hbar)    \Big)   \,,\\
&\widetilde\W_{\T[Q_K] }  ( \mathbf{x} ,\mathbf{y}  )   =\sum\limits_i \Li_2(y_i) +  \log\,((-1)^{C_{ii}} x_i  )\, \log \, y_i  + \sum\limits_{i,j}   \frac{C_{ij} }{2} \log \, y_i \, \log \, y_j \,.\label{superpotentialknot}
\end{align}
By comparing \eqref{superpotentialknot} with \eqref{WeffQ}, we note that the lifted HOMFLY-PT polynomials $P^{Q_K}(\mathbf{x},q)$ are the same as vortex partition functions of $\T_{A,N}$ theories, and the corresponding quiver theories $T[Q_K]$ are actually 
\begin{align}
\T_{A,N}:~~(U(1)-[1])^{  \otimes N}_{k_{ij},~\xi_i}\,.
\end{align}
Therefore $C_{ij}$ play the role of effective Chern-Simons levels $k_{ij}^{eff}$ and $\log\,((-1)^{C_{ii}} x_i  )$ play the role of FI parameters $\xi_{i}^{eff}$. 
The mirror transformations of $\T_{A,N}$ theories enable us to obtain a chain of equivalent integer matrices $\{C_{ij}\}$ for knots.

\subsection*{Trefoil.}
Trefoil $\mathbf{3_1}$ is one typical example in KQ correspondence \cite{Kucharski:2017ogk,Ekholm:2018eee}. The associated KQ matrix $C_{ij}$ is 
\begin{align}
C_{ij}=
\left(
\begin{array}{c c c c c c}
0 &~ 0 &~ 1 &~1 &~2 &~2 \\
0 & 1 & 1 &1& 2&2  \\
1 &  1 & 2 &2 &2 &3\\
1 & 1 & 2 &3 & 2 &3\\
2 &2 &2 &2 &3 &3\\
2 & 2 &3 &3 &3 &4
\end{array}
\right)
+
f
\left(
\begin{array}{c c c c c c}
1 &~ 1 &~ 1 &~1 &~1 &~1 \\
1 & 1 & 1 &1& 1&1  \\
1 &  1 & 1 &1 &1 &1\\
1 & 1 & 1 &1 & 1 &1\\
1 &1 &1 &1 &1 &1\\
1 & 1 &1 &1 &1 &1
\end{array}
\right)
\end{align}
where $f$  in the second term is the framing number for trefoil. 
Based on the above conjecture that the 3d theory $\T[Q_K]$ from KQ correspondence is the $\T_{A,N }$ theory,
we assume the original theory denoted by $\T[  \mathbf{(0, 0,0,0,0,0)}]$ has
effective CS levels
\begin{align}
C_{ij}=k_{ij}^{eff, \, \mathbf{( 0, \dots, 0)}}=k_{ij} +\frac{1}{2} \delta_{ij} \,,
\end{align}
 and  mass parameters were absorbed into shifted FI parameters $\txi_i$. 
 %The sphere partition function is provided by \eqref{partfuncT_{AN}}.
 Then one can act with mirror  transformations from $\H(\T_{A,6} )$ on the sphere partition function given in  \eqref{partfuncT_{AN}} and get many integer effective CS level matrices.
 %Interested readers can contact authors for the calculation data. 
 
\begin{figure}
	\centering
	\includegraphics[width=1.5in]{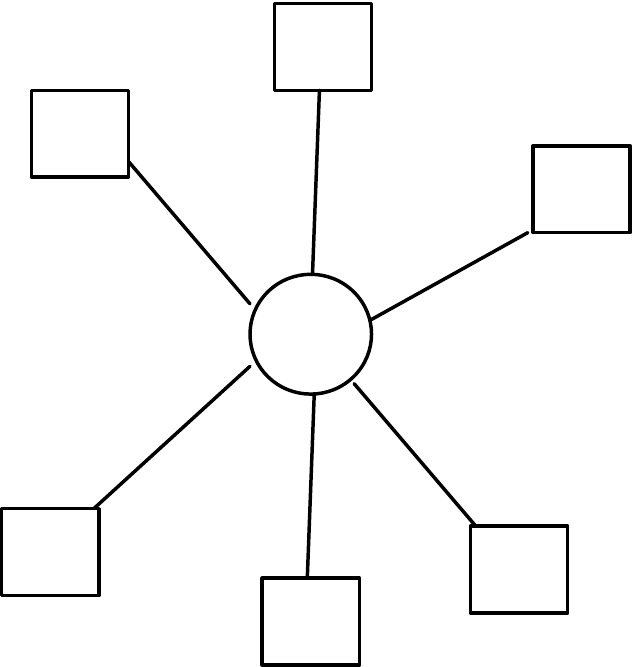} 
	\caption{The star shape quiver for the 3d $\N=2$ theories corresponding to trefoil.}
	\label{fig:star}
\end{figure}

 Quiver reductions appear in this context as well. By scanning the CS levels obtained by mirror transformations, we find there is at least one gauge node that cannot be integrated out. More explicitly, mirror transformation $\mathbf{(0,1,1,1,1,1 )}$ leads to the sphere partition function
\begin{align}
&Z_{S_b^3}^{\T[ \mathbf{ ( 0,1,1,1,1,1 ) }]  } = \nn\\
&\int  dx\,e^{ - \frac{1}{2} ( 9+14 f+5 f^2) \pi i \, x^2 +
	\pi i \, x \big( -\frac{i (6+5f)}{4} Q+
	( 2 \txi_1+f \txi_2 +(1+f) \txi_3 +(1+f) \txi_4 + (2+f) \txi_5  +(2+f) \txi_6) 
	\big)   }    \nn\\
& \quad \times
s_b\big( \frac{i Q}{2} -x \big)
s_b\big( \frac{i Q}{4} +f\, x-\txi_2 \big)
s_b\big( \frac{i Q}{4} +(1+f)\,x  -\txi_3 \big)
s_b\big( \frac{i Q}{4} +(1+f)\, x  -\txi_4 \big)  \nn\\
&\quad \times
s_b\big( \frac{i Q}{4} +(2+f)\,x  -\txi_5 \big)
s_b\big( \frac{i Q}{4} +(2+f)\,x  -\txi_6 \big) \,,
\end{align}
which implies that the corresponding theory has a star shape quiver in figure \ref{fig:star} with one gauge node $U(1)$ and six chiral multiplets with charges $\{ -1, f, 1+f,1+f,2+f , 2+f \}$. The FI parameters $\txi_{2,3,4,5,6}$ were turned into mass parameters while $\txi_1$ is still an FI parameter. 
If $f=0, -1, -2$, some double sine functions from chiral multiplets can be moved out of the integral, so framing $f$ plays a subtle role here. Moreover, mirror transformation $\mathbf{ (1,1,1,1,0,1 )}$ also leads to a star shape quiver with one gauge node $U(1)$ and six chiral multiplets with charges $\{ 2+f, 2+f, 2+f , 2+f, -1, 3+f \}$. The corresponding sphere partition function is
\begin{align}
&Z_{S_b^3}^{\T[ \mathbf{ (1,1,1,1,0,1 ) }]  } = \nn\\
& \int  dx\,e^{ - \frac{1}{2} ( 30+24 f+5 f^2)\pi i \, x^2 +
	\pi i \, x 
	\big( -\frac{11+5 f }{4} i\, Q+
	( 2 \txi_1+2 \txi_2 +2 \txi_3 +2 \txi_4 + 2 \txi_5  +3 \txi_6) +f (\txi_1+\txi_2 +\txi_3+\txi_4+\txi_6 )
	\big)   }   \nn\\
& \qquad \times
s_b\big( \frac{i Q}{4} +(2+f)x-\txi_1 \big)
s_b\big( \frac{i Q}{4} + (2+f)\, x-\txi_2 \big)
s_b\big( \frac{i Q}{4} +(2+f)\,x  -\txi_3 \big)
 \nn\\
& \qquad \times
s_b\big( \frac{i Q}{4} +(2+f)\, x  -\txi_4 \big) 
s_b\big( \frac{i Q}{2} -\,x  -\txi_5 \big)
s_b\big( \frac{i Q}{4} +(3+f)\,x  -\txi_6 \big) \,.
\end{align}
In this case, all FI parameters $\txi_{1,2,3,4,5,6}$ are turned into mass parameters.

\section{Conclusion}\label{sec-5}

In this work we discussed the mirror symmetry for abelian 3d $\N=2$ gauge theories using  $\T_{A,N}$ theories, on which mirror symmetry acts as a functional Fourier transformation of sphere partition functions. These transformations form a nice group $\H(\T_{A,N})$, so that each element in $\H( \T_{A,N})$ stands for a mirror transformation and corresponds to a mirror dual theory. By reading off effective mixed Chern-Simons levels and effective FI parameters from superpotentials, we can get many mirror dual theories with different mixed CS levels. However, these mirror dual theories are equivalent and have equivalent partition functions. 
This implies that effective CS levels are not sufficient to identify theories in this context. Fortunately, these equivalent mixed CS levels can be tracked by mirror transformations and are under control. As many theories are related to $ \T_{A,N}$ theory, and the latter theory is easy to analyze, we can use $\T_{A,N}$ as a tool to analyze other types of quiver theories.
We discussed the 3d mirror symmetry of theories engineered by strip geometries, in particular
$U(1)_k-[N]$ theories, by turning them into $\T_{A,N}$ theories via mirror transformation $\mathbf{(1,1,\dots, 1)}$. 
%One can then perform mirror transformations on its equivalent $\T_{A,N}$ theory.
The result is that for these theories there are several corresponding mirror dual $\T_{A,N}$ theories with different mixed CS level matrices. If considering their vortex partition functions, 
%We also obtain the same results from vortex partition functions. 
one could find mirror symmetry only changes the sign of mass parameters.
An interesting discovery is that Tong's mirror pairs can be verified with the help of $\T_{A,N}$ theories.
In addition, we discussed the open BPS invariants encoded in vortex partition functions, and the open Gopakumar-Vafa formula in various limits.

There are many open questions.
First, it would be interesting to understand quiver reductions, and the relations between mixed CS levels and charge vectors for chiral multiplets.  Second, it is important to understand better
mirror transformations and quiver reductions for knot polynomials and their Higgsing and geometric realization.  Third, finding the relations between non-abelian 3d $\N=2$ theories with mixed Chern-Simons levels, 3d/3d correspondence, three-manifolds, cluster algebra, superpotentials and monopole operators, is an interesting direction for further studies \cite{Terashima:2014aa,Gang:2015aa}.  %Since mirror transformations can be applied not only to 3d $\N=2$ theories but also 3d $\N=4$ theories, it is interesting to see if there are mirror equivalent CS levels for the later.
 Last but not the least, it is important to verify whether the local mirror symmetry discussed in \cite{Collinucci:2016aa} can be identified with the mirror symmetry discussed in this work, and find the mirror symmetry for 3d $\N=2$ theories obtained by compactifying 6d $(2,0)$ SCFTs on three manifold $M_3$. 
 % see \cite{Dimofte:2011ju, Closset:2018ghr,Eckhard:2019aa,Closset:2017zgf,Fan:2020aa}  for recent development.

%====================================================================
%\input{sec-conclusion.tex}
%====================================================================
\acknowledgments
We would like to thank particularly Piotr Su\l{}kowski for helping to improve this paper.  We also thank Mohammad Akhond and Masahito Yamazaki for helpful comments.
 The work of S.C. is supported by 
TEAM program of the Foundation for Polish Science cofinanced by the European Union
under the European Regional Development Fund (POIR.04.04.00-00-5C55/17-00).

\bigskip

\appendix 

\section{Double-sine function}

The double-sine function is defined as 
\begin{align}
s_b(x)=\prod\limits_{m,n\geqslant 0} \frac{ m\, b +n/b+ Q/2- i\, x   }{ m\, b +n/b+Q/2 +i\, x    }\,,~~ ~~Q=b+\frac{1}{b}
\end{align}
and it satisfies the identity
\begin{align}
s_b(x)\,s_b(-x)=1 \,.
\end{align}
The equivariant parameter $q$ in localization is defined as
\begin{align}\label{Qqh}
q=e^{\hbar}=e^{ 2\pi \, i \, b^2 }=e^{2 \pi  \, i \, b \,Q}\,,~~ \hbar=2\pi \, i \, b^2   =2 \pi  \, i \, b \,Q   \,.
\end{align}
The asymptotic limit $b\rightarrow 0$ of the double-sine function is 
\begin{align}\label{sbasym}
s_b(z)\rightarrow e^{ - i\, \pi  \, z^2/2 } e^{  i \, \pi (2-Q^2)/24} \exp \left(  \frac{1}{2 \pi \, i \, b^2}     \Li_2( e^{2 \pi b z  }  ) \right) 
\end{align}
where $\Li_2(z)$ is the polylogarithm function defined by a power series 
\begin{align}
\Li_s(z):=\sum\limits_{k=1}^{\inf} \frac{z^k}{k^s}  \,.
\end{align}
In the decompactification limit $R\rightarrow +\inf$, the effective superpotentials of 3d $\N=2$ gauge theories on spacetime $\mathbb{R}^2\times S_R^1$ involve
\begin{align}\label{limitLi2}
\underset{R\rightarrow +\inf}{\text{lim}} \frac {\Li_2 \big(  e^{-R\, x} \big) }{R^2 }   = \frac{    \text{\textlbrackdbl}  x \text{\textrbrackdbl}   ^2   }{2}  \,, \qquad
\text{\textlbrackdbl}  x \text{\textrbrackdbl}   ^2 
:=  \theta(-x) \cdot x 
= \left\{ \begin{array}{cc}
0 &\quad x >0\, , \\
x &\quad x < 0\, , 
\end{array}  \right.  
\end{align}
where $  \text{\textlbrackdbl}  x \text{\textrbrackdbl}   ^2  $ is defined in \cite{Hayashi:2019aa} and $\theta(x)$ is the Heaviside step function.
The derivative of $\Li_2(y)$ in vacua equations is
\begin{align}
\exp \Big({   y \frac{ d\, \Li_2 (y) }{  d\, y} }\Big)=\frac{1}{ 1-y}  \,.
\end{align}
There is one useful identity in reading off effective superpotentials
\begin{align}
\Li_2(z)+\Li_2(z^{-1})=-\frac{\pi^2 }{6 } -\frac{1}{2} \, \log^2(-z)     \,.
\end{align}
In addition,  the $q$-Pochhammers is defined by $(x;q)_n:= \prod_{i=0}^{n-1}(1-x q^i) $.

\section{ Integration}
When performing mirror transformations, we use the higher dimensional Gaussian integral formula
\begin{align}\label{integralAJ}
\int \text{d}\, \mathbf{x} \, \exp  \Big(-\frac{1}{2}  \,\textbf{x} \cdot \textbf{A}\cdot \textbf{x}  + \textbf{J}\cdot \textbf{x} \Big )\,  = \sqrt{ \frac{(2 \pi)^n  }{ \text{det} \textbf{A}  }  } \,\exp \Big(   \frac{1}{2} \, \textbf{J} \cdot \textbf{A}^{-1} \cdot \textbf{J}  \Big) \,,~~\text{only if ~}\text{det} \textbf{A}  \neq 0  \,,
\end{align}
to integrate out old gauge nodes.
The Dirac delta function 
\begin{align}
\delta(k)=\frac{1}{2 \pi} \int d x\, e^{ i\, k x} 
\end{align}
reduces the dimension of integrals and hence plays an important role in quiver reduction.

\section{ Matrix decomposition}
Real symmetric matrix $\bf{S}$ can be decomposed in the orthogonal basis
\begin{align}
\textbf{S}=\mathbf{Q}^T \mathbf{\Lambda} \mathbf{Q} \,,
\end{align}
where $\bf{\Lambda}$ is a real diagonal matrix and $\mathbf{Q}$ is an orthogonal matrix satisfying $\mathbf {Q} \,\mathbf{Q}^T=\mathbf {Q}^T \mathbf{Q}=1\,,\text{and~} \mathbf{Q}^T=\mathbf{Q}^{-1}$.
If matrix $\bf{A}$ is symmetric, then $\mathbf{B}^T \mathbf{A} \mathbf{B}$ and $ \mathbf{A}^{-1}$ are also symmetric.
In addition, Cholesky decomposition asserts that if matrix $\bf A$ is real positive and definite symmetric, then  it can be decomposed as 
$
\mathbf{ A}= \mathbf{L} \,\mathbf{L}^T   
$,
or more specifically
$
\textbf{A}_{ik}= \textbf{L}_{ij} \textbf{L}_{kj}   $.

%%%%%%%%%%%%%%%%%%%%%%%%%%%%%%%%%%%%
\bibliographystyle{JHEP}
\bibliography{ref}

\end{document}